%% file: main.tex
\documentclass[runningheads]{llncs}
\usepackage[T1]{fontenc}
\usepackage{tikz}
\usepackage{array}
\usepackage{units}
\usepackage{bm}
\usepackage{amssymb}
\usepackage[nolist,nohyperlinks]{acronym}
\usepackage[hidelinks]{hyperref}
\usepackage{wrapfig}
\usetikzlibrary{positioning,calc,arrows,shapes,automata}

\usepackage{color}

\hypersetup{
	colorlinks = true,
	linkcolor=blue,
	citecolor=blue
}

\newcolumntype{P}[1]{>{\centering\arraybackslash}p{#1}}

\begin{acronym}
	
	\acro{sul}[SUL]{System Under Learning}
	\acroplural{sul}[SULs]{Systems Under Learning}
	
	\acro{dfa}[DFA]{Deterministic Finite Automaton}
	
	\acro{mat}[MAT]{Minimally Adequate Teacher}
	
	\acro{ble}[BLE]{Bluetooth Low Energy}
	
	\acro{iot}[IoT]{Internet of Things}
	
	\acro{mtu}[MTU]{maximum transmission unit}
	
	\acro{soc}[SoC]{System on a Chip}
	\acroplural{soc}[SoCs]{Systems on a Chip}
	
	\acro{mqtt}[MQTT]{Message Queuing Telemetry Transport}
	
	\acro{gatt}[GATT]{Generic Attribute Profile}
	
	\acro{ll}[LL]{Link Layer}

	\acro{sm}[SM]{Security Manager}
	
\end{acronym}

\newcommand{\new}[1]{#1}

\def\circledarrow#1#2#3{ 
		\draw[#1,->] (#2) +(80:#3) arc(80:-260:#3);
	}

\begin{document}
	
	\title{Fingerprinting and Analysis of Bluetooth Devices with Automata Learning}

	\author{Andrea Pferscher\orcidID{0000-0002-9446-9541} \and \\
	Bernhard K.~Aichernig\orcidID{0000-0002-3484-5584}}

\authorrunning{Pferscher and Aichernig}

\institute{Institute of Software Technology, Graz University of Technology, Graz, Austria \\
	\email{\{apfersch,aichernig\}@ist.tugraz.at}}
	
	\maketitle
	
	\begin{abstract}
		Automata learning is a technique to automatically infer behavioral models of black-box systems. Today's learning algorithms enable the deduction of models that describe complex system properties, e.g., timed or stochastic behavior. Despite recent improvements in the scalability of learning algorithms, their practical applicability is still an open issue. Little work exists that actually learns models of physical black-box systems. To fill this gap in the literature, we present a case study on applying automata learning on the Bluetooth Low Energy (BLE) protocol. It shows that not \new{only} the size of the system limits the applicability of automata learning. 
	\new{Also}, the interaction with the system under learning \new{creates} a major bottleneck that is rarely discussed. In this article, we propose a general automata learning architecture for learning a behavioral model of the BLE protocol implemented by a physical device. With this framework, we can successfully learn the behavior of six investigated BLE devices. Furthermore, we extended the learning technique to learn security critical behavior, e.g., key-exchange procedures for encrypted communication. The learned models depict several behavioral differences and inconsistencies to the BLE specification. This shows that automata learning can be used for fingerprinting black-box devices, i.e., \new{characterizing} systems via their specific learned models. Moreover, learning revealed a crashing scenario for one device.


	\end{abstract}

	\section{Introduction}\label{sec:intro}
	
	\input{intro}
	
	\section{Preliminaries} \label{sec:preliminaries}
	
	\input{prelim}

	\section{Learning Setup}\label{sec:learning-setup}
	
	\input{architecture}

	\section{Evaluation}\label{sec:evaluation}
	
	\input{case-study}

	\section{Related Work} \label{sec:related-work}
	
	\input{related-work}

	\section{Conclusion}\label{sec:conclusion}
	
	\input{conclusion}

	\section*{Acknowledgement}
	This work is funded by the TU Graz LEAD project \emph{Dependable Internet of Things in Adverse Environments}, by the \emph{LearnTwins} project (No 880852) from the Austrian Research Promotion Agency (FFG), and by \emph{AIDOaRt} project (grant agreement No 101007350) from the ECSEL Joint Undertaking (JU). The JU receives support from the European Union’s Horizon 2020 research and innovation programme and Sweden, Austria, Czech Republic, Finland, France, Italy, and Spain. 
	We would like to thank Maximilian Schuh for providing support for the BLE devices and the authors of the SweynTooth paper for creating an open-source BLE interface. This version of the article has been accepted for publication, after peer review but is not the Version of Record and does not reflect post-acceptance improvements, or any corrections. The Version of Record is available online at: \url{https://doi.org/10.1007/s10703-023-00425-y}.
	
	\bibliographystyle{splncs04}
	\bibliography{main}
	
\end{document}

%% file: intro.tex
Bluetooth is a key communication technology in many different fields. Currently, it is assumed that \new{$4.7$} billion Bluetooth devices are shipped annually and that the number will grow to \new{seven} billion by 2026 \cite{bleMarket}. This growth mainly refers to the increase of peripheral devices that support \ac{ble}. With \ac{ble}, Bluetooth became also accessible for low-energy devices. Hence, \ac{ble} is a vital technology in the \ac{iot}.

The amount of heterogeneous devices in the \ac{iot} makes the assurance of dependability a challenging task, especially, since the insight into \ac{iot} components is frequently limited. \new{For example, Texas Instruments~\cite{texasInstruments} motivates in a technical report that wired communication in a car can be replaced by \ac{ble}. Considering that automotive components are developed by many different suppliers, the used \ac{ble} chip and, more likely, the installed firmware version might be unknown. Facing such challenges}, the system under test must be considered a black box.

Enabling in-depth testing of black-box systems is difficult, but can be achieved with model-based testing techniques. Garbelini et al.~\cite{DBLP:conf/usenix/Garbelini00SK20} successfully used a generic model of the \ac{ble} protocol to detect security vulnerabilities of \ac{ble} devices via model-based fuzzing. However, they state that the creation of such a comprehensive model was challenging since the \ac{ble} protocol is underspecified.

To overcome the possibly tedious and error-prone process of model creation, learning-based testing techniques have been proposed \cite{DBLP:conf/dagstuhl/AichernigMMTT16}. Learning-based testing applies automata learning algorithms to automatically infer a behavioral model of a black-box system. The learned model could then be used for further verification \new{and testing}. 

\new{Existing work \cite{DBLP:conf/uss/RuiterP15,DBLP:conf/cav/Fiterau-Brostean16,DBLP:conf/spin/Fiterau-Brostean17,DBLP:conf/icst/TapplerAB17,DBLP:conf/esorics/StoneCR18,DBLP:conf/uss/Fiterau-Brostean20} applied learning-based testing to create behavioral models of communication protocols like TLS, TCP or SSH. The learned models show behavioral inconsistencies to the specification and security vulnerabilities. In the literature, this technique is also known as protocol state fuzzing.}

Motivated by promising results of protocol state fuzzing, various automata learning algorithms have been proposed to extend learning for more complex system properties like timed \cite{DBLP:conf/formats/TapplerALL19,DBLP:conf/nfm/AichernigPT20} or stochastic behavior \cite{DBLP:conf/fm/TapplerA0EL19}. However, few evaluations of these algorithms on implementations on physical devices exist.

\begin{figure}[t]
	\centering
	\includegraphics{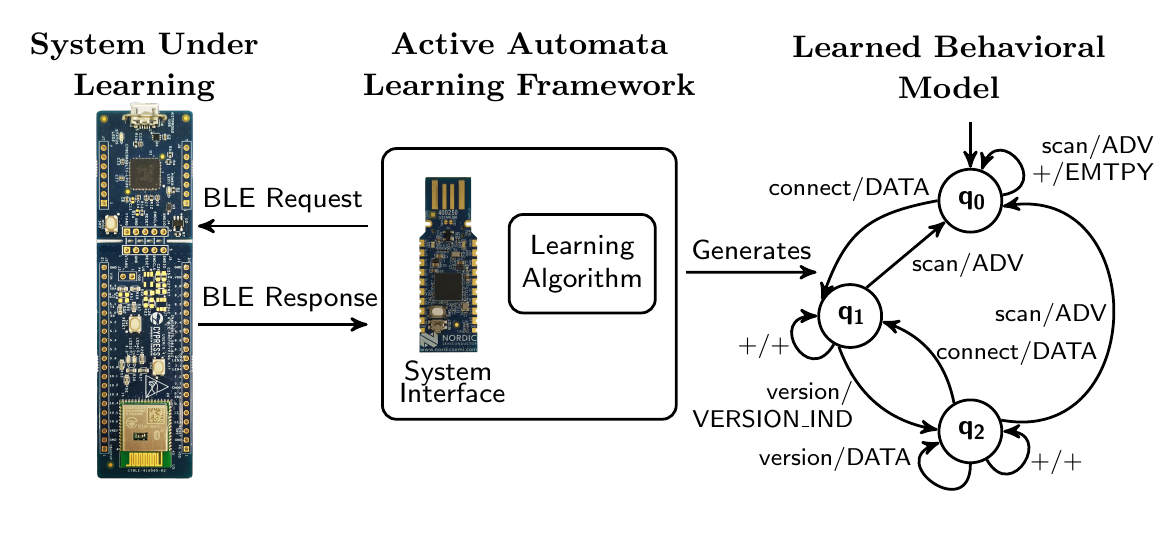}
	\caption{Automata learning framework for the inference of behavioral models of \ac{ble} devices. The model is generated via an active interaction through a system interface that includes an additional \ac{ble} device for communication.}
	\label{fig:framework-overview}
\end{figure}

In this article, we present a case study that applies automata learning on \ac{ble} devices. Figure~\ref{fig:framework-overview} illustrates the basic concept presented in this article. Our objective is to learn the behavioral model of the \ac{ble} protocol implementation on a physical device. For this, we propose a general automata-learning framework that automatically infers the behavioral model of \ac{ble} devices. Our presented framework uses state-of-the-art automata learning techniques. The learning algorithm actively queries the \ac{sul}. To enable an active interaction with the \ac{sul}, we propose a system interface that communicates via \ac{ble} with the \ac{sul}. 
\new{For this, we include in our interface an additional \ac{ble} device that enables the sending of custom \ac{ble} packets. The system interface also allows handling encrypted communication.} Furthermore, we adapt the learning algorithm considering practical challenges that occur in learning real network components.


In our case study, we present our results on learning six different \ac{ble} devices. For all six devices, we learn the behavior during the \ac{ble} connection establishment. In addition, for three out of these six devices, we learn a model of the \ac{ble} pairing process. The pairing process includes security critical behavior like the exchange of keys to establish an encrypted communication. Based on our results, we discuss three different findings. First, we observe that the implementations of the \ac{ble} stacks differ from device to device.
Using this observation, we show that active automata learning can be used to \new{fingerprint} black-box systems. Second, the presented performance metrics show that not only does the system's size influences the performance of the learning algorithm, but also the creation of a deterministic learning setup creates a significant overhead which has an impact on the efficiency of the learning algorithm since we have to repeat queries and wait for answers. 
Third, learning reveals robustness issues of the investigated devices. We observe that devices might crash upon unexpected input sequences.


The paper includes the following contributions: 

\begin{enumerate}
	\item A learning framework that enables learning of \ac{ble} protocol implementations of peripheral devices. The framework including the learned models are available \new{online}~\cite{bleLearning}.
	\item An extension of the learning framework that enables the learning of the key-exchange protocol during the \ac{ble} pairing procedure.
	\item A case study that evaluates our framework on real physical devices.
	\item \new{A model-based manual analysis that checks if the devices conform to the \ac{ble} specification.} 
	\item A sequence found by learning that crashes a \ac{ble} device. 
	\item \new{A sequence that allows fingerprinting of the investigated \ac{ble} devices.} 
\end{enumerate}

\new{One future purpose of our work is to develop model-based testing techniques for \ac{ble} using our learned models. Following the model-based fuzzing technique by Garbelini et al.~\cite{DBLP:conf/usenix/Garbelini00SK20}, in the next step, our learning framework could facilitate such a black-box fuzzer by automatically learning the model. In our follow-up work~\cite{DBLP:conf/nfm/PferscherA22}, we have already applied the learning framework that we propose in this article to generate a stateful black fuzzer for \ac{ble}. Our learning-based fuzzing technique revealed inconsistencies to the \ac{ble} specification and robustness issues in the investigated \ac{ble} devices.}


This article is an extended version of our conference paper ``Fingerprinting Bluetooth Low Energy Devices via Active Automata Learning'' \cite{DBLP:conf/fm/PferscherA21} presented at \emph{Formal Methods - 24th International Symposium}. 
Additional content and contributions in this preprint article include the extension of our learning framework for the \ac{ble} pairing procedure. \new{For this, we create an advanced learning interface that enables the establishment of encrypted communication with the peripheral. Further extensions, include an advanced caching technique that deals with non-deterministic behavior to increase the robustness of our proposed learning technique. Furthermore, we consider an additional \ac{ble} device in the case study and the identification of a crashing sequence for a \ac{ble} device.}

The article is structured as follows. Section~\ref{sec:preliminaries} discusses the used modeling formalism, active automata learning, and the \ac{ble} protocol. In Section~\ref{sec:learning-setup}, we propose our learning architecture, followed by the performed evaluation based on this framework in Section~\ref{sec:evaluation}. Section~\ref{sec:related-work} discusses related work and Section~\ref{sec:conclusion} concludes the article. 

%% file: prelim.tex
\subsection{Mealy Machines}

Mealy machines represent a neat modeling formalism for systems that create observable outputs after an input execution, i.e., reactive systems.  Moreover, many state-of-the-art automata learning algorithms and frameworks \cite{DBLP:conf/cav/IsbernerHS15,aalpy} support Mealy machines. A Mealy machine is a finite state machine, where the states are connected via transitions that are labeled with input actions and the corresponding observable outputs. Starting from an initial state, input sequences can be executed and the corresponding output sequence is returned. 
\begin{definition}[Mealy machine]\label{def:mealy-machines}
	A Mealy machine is a 6-tuple $\mathcal{M} = \langle Q, q_0, I, $ $ O, \delta, \lambda \rangle$ where 
	\begin{itemize}
		\item $Q$ is the finite set of states
		\item $q_0$ is the initial state
		\item $I$ is the finite set of inputs
		\item $O$ is the finite set of outputs
		\item $\delta \colon Q \times I \rightarrow Q$ is the state-transition function
		\item $\lambda \colon Q \times I \rightarrow O$ is the output function
	\end{itemize}
\end{definition}
To ensure learnability, we require $\mathcal{M}$ to be deterministic and input-enabled. Hence, $\delta$ and $\lambda$ are total functions.
Let $S$ be the set of observable sequences, where a sequence $s \in S$  consists of consecutive input/output pairs $(i_1,o_1),\ldots, $ $(i_j,o_j),\ldots, (i_{n},o_{n})$ with $i_j \in I$, $o_j \in O$, $j \leq n$ and $n \in \mathbb{N}$ defining the length of the sequence. We define $s_I \in I^*$ as the corresponding input sequence of $s$, and $s_O \in O^*$ maps to the output sequence. We extend $\delta$ and $\lambda$ for sequences. The state transition function $\delta^* : Q \times I^* \rightarrow Q$ gives the reached state after the execution of the input sequence and the output function $\lambda^* : Q \times I^* \rightarrow O^*$ returns the observed output sequence. 
We define two Mealy machines $\mathcal{M} = \langle Q, q_0, I, O, \delta, \lambda \rangle$ and $\mathcal{M}' = \langle Q', q_0', I, O, \delta', \lambda' \rangle$ as \new{equivalent if there exists no $s_I \in I^*$ such that $\lambda^*(q_0, s_I) \neq \lambda'^*(q_0',s_I)$}, i.e., the execution of all input sequences leads \new{to the same} output sequences. 

\subsection{Active Automata Learning}
In automata learning, we learn a behavioral model of a system based on a set of execution traces. Depending on the generation of these traces, we distinguish between two techniques: \emph{passive} and \emph{active} learning. Passive techniques reconstruct the behavioral model from a given set of traces, e.g., log files.
\new{Behavior that is not covered in the data set, can only be approximated by generalizations. Hence, incomplete data presents a problem for passive learning since the generalization for unusual behavior could be inadequate. For example, ordinary log files might not cover the behavior of unusual input sequences.
In our previous work \cite{DBLP:journals/corr/abs-2209-14031}, we showed that passive learning requires a significantly larger data set than active learning to cover the same behavior with randomly generated input sequences. Therefore, the attempt to include rare behavior via random sequences requires a large number of samples.} Active techniques, instead, actively query the \ac{sul}. As a result, actively learned models are more likely to cover rare events that cannot be observed from ordinary system monitoring.

\begin{figure}[t]
	\centering
	\input{mat-framework.tex}
	\caption{\new{An adapted version of \acf{mat} framework proposed by Angluin \cite{DBLP:journals/iandc/Angluin87}. The adaptions include the learning of Mealy machines, the replacement of the equivalence oracle by conformance testing, and the concept of abstraction by a mapper component. This figure has been adapted from the illustration proposed by Aichernig et al.~\cite{DBLP:conf/dagstuhl/AichernigMMTT16}.}}
	\label{fig:mat-framework}
\end{figure}
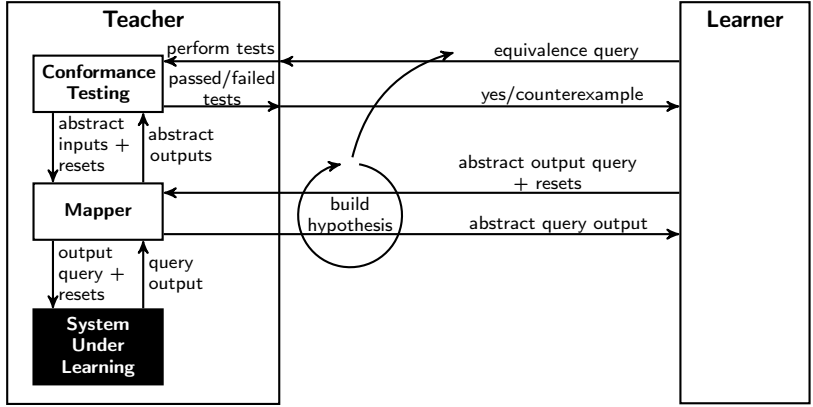

Many current active learning algorithms build upon the $L^*$ algorithm proposed by Angulin~\cite{DBLP:journals/iandc/Angluin87}. The original algorithm learns the minimal \ac{dfa} of a regular language.
Angluin's seminal work introduces the \acf{mat} framework \new{which we illustrate in Figure~\ref{fig:mat-framework}.} The framework comprises two members: the \emph{learner} and the \emph{teacher}. The learner constructs a \ac{dfa} by questioning the teacher, who knows the \ac{sul}. The \ac{mat} framework distinguishes between \emph{membership} and \emph{equivalence} queries. Using membership queries, the learner asks if a word is part of the language, which can be either answered with \emph{yes} or \emph{no} by the teacher. Based on these answers, the learner constructs an initial behavioral model. The constructed hypothesis is then provided to the teacher to ask if the \ac{dfa} conforms to the \ac{sul}, i.e., the learner queries equivalence. The teacher answers equivalence queries either with a counterexample that shows non-conformance between the hypothesis and the \ac{sul} or by responding \emph{yes} to affirm conformance. In the case that a counterexample is returned, the learner uses this counterexample to pose new membership queries and construct a new hypothesis. This iterative procedure is repeated until a conforming hypothesis is proposed. 

The $L^*$ algorithm has been extended to learn Mealy machines of reactive systems \cite{DBLP:conf/hldvt/MargariaNRS04,DBLP:phd/de/Niese2003,DBLP:conf/fm/ShahbazG09}. \new{Figure~\ref{fig:mat-framework} includes the adaptions for learning Mealy machines, where membership queries are replaced by output queries.} For this, the learner asks for the output sequence produced by a given input sequence. We assume that the teacher has access to the \ac{sul} to execute inputs and observe outputs. \new{Furthermore, Angluin's $L^*$ algorithm requires the \ac{sul} to be resettable to an initial state.}  

In practice, we cannot assume a \emph{perfect} teacher who provides the counterexample that shows non-conformance between the hypothesis and the \ac{sul}. To overcome this problem, we use conformance testing to substitute equivalence queries. \new{According to the definition of Lee and Yannakakis~\cite{DBLP:journals/pieee/LeeY96}, conformance testing assesses if an implementation conforms to a specification. They assume that the specification is a Mealy machine and the implementation is a black box where outputs to the corresponding input sequences are observable. For learning, we want to test if our learned model correctly defines the behavior of our \ac{sul}. For this, we test the conformance between the \ac{sul} and the provided hypothesis.}   

\new{We assume that the behavior of the \ac{sul} can be represented by a Mealy machine. For this, we can define the conformance relation based on the equivalence of Mealy machines. However, since the final number of states of the \ac{sul} is unknown, we can only approximate conformance by a set of finite input/output sequences. For this, we say that the learned hypothesis $\mathcal{H} = \langle Q, q_0, I, O, \delta, \lambda \rangle$  conforms to the \ac{sul} $\mathcal{I} = \langle Q', q'_0, I', O', \delta', \lambda' \rangle$ if for a finite set of input sequences $\mathcal{S}_I$ the following relation holds.}

\begin{equation}\label{eq:conformance-relation}
	\forall s_I \in \mathcal{S}_I \colon \lambda^*(q_0, s_I) = \lambda'^*(q_0',s_I)
\end{equation}

\new{The goal in conformance testing during learning is to find an input sequence that violates this conformance relation. In the case a counterexample  is found, the teacher provides such an input sequence to the learner as a counterexample to the conformance between the hypothesis and the \ac{sul}. The learner uses this counterexample to refine the hypothesis by performing further output queries. The refinement of the hypothesis is repeated until no counterexample to the conformance between the hypothesis and the \ac{sul} can be found.}

\new{The complexity of automata learning depends on the number of states and the considered input alphabet. Especially, in the learning of communication protocols, considering all possible inputs would make learning infeasible. Cho et al.~\cite{DBLP:conf/ccs/ChocSS10} introduce the concept of abstraction to make the learning of communication protocols feasible. For this, instead of considering a large set of inputs, a smaller set of abstract inputs and outputs is considered for learning. Hence, the learned model represents an abstraction of the \ac{sul}. Aarts et al.~\cite{DBLP:journals/fmsd/AartsJUV15} present the concept of abstraction with the introduction of a mapper component. The purpose of the mapper is to translate the input and output actions respectively. For this, abstract inputs for learning are translated by the mapper into concrete inputs that can be executed on the \ac{sul}. For outputs, the mapper translates the received concrete outputs into abstract outputs. Figure~\ref{fig:mat-framework} illustrates the placement of the mapper in the \ac{mat} framework.}

\subsection{Bluetooth Low Energy} \label{sec:ble}


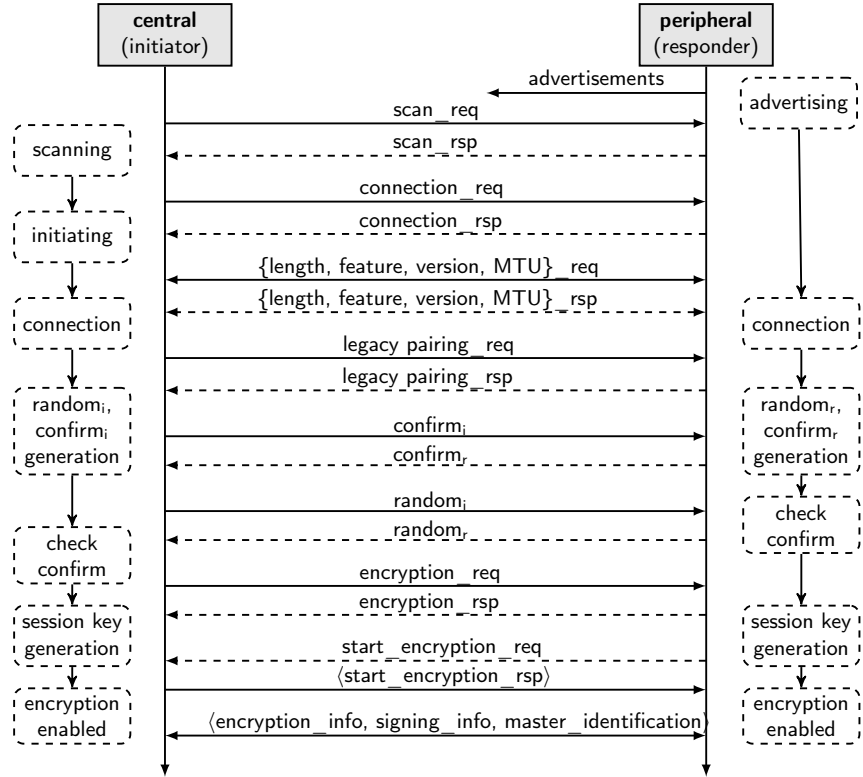
\begin{figure}[t]
	\centering
	\scalebox{0.9}{
		\input{ble-sequence-diagram}
	}
	\caption{Communication between a BLE central and peripheral device to establish a connection. After the connection is established, the pairing procedure starts. In the legacy pairing method, the central and peripheral exchange values to generate a session key that is used for an encrypted communication (indicated by $\langle \ldots \rangle$).  The sequence diagram is taken from Garbelini et al.~\cite{DBLP:conf/usenix/Garbelini00SK20} and extended by the message sequence for the legacy pairing procedure taken from the \ac{ble} specification \cite{Bluetooth53}.}
	\label{fig:ble-communication} 
\end{figure}

The \ac{ble} protocol is a lightweight alternative to the classic Bluetooth protocol, specially designed to provide a low-energy alternative for \ac{iot} devices.  The Bluetooth specification \cite{Bluetooth53} defines the connection and pairing protocol between two \ac{ble} devices according to different layers of the \ac{ble} protocol stack. Figure~\ref{fig:ble-communication} shows the initial communication messages of two \ac{ble} devices that first establish a connection and then exchange parameters to establish an encrypted communication. We distinguish between the \emph{peripheral} and the \emph{central} device. \new{An example of central device would be smartphone that wants to connect with a peripheral device, e.g., a smart watch.} In the remainder of this article, we refer to the central device simply as \emph{central} and to the peripheral device as \emph{peripheral}. 

The peripheral sends advertisements to show that it is available for connection with a central. According to the \ac{ble} specification, the peripheral is in the \emph{advertising} state.
If the central scans for advertising devices in the \emph{scanning} state. For this, the central sends a scan request ($\mathsf{scan\_req}$) to the peripheral, which responses with a scan response ($\mathsf{scan\_rsp}$). In the next step, the central changes from the \emph{scanning} to the \emph{initiating} state by sending the connection request ($\mathsf{connection\_req}$). If the peripheral answers with a connection response ($\mathsf{connection\_rsp}$), the peripheral and central enter the \emph{connection} state. After the connection, the negotiation on communication parameters starts. Both, the central and peripheral can request features or send control packets. These request and control packets include maximum packet length, \ac{mtu}, \ac{ble} version, and  feature exchanges. As noted by Garbelini et al.~\cite{DBLP:conf/usenix/Garbelini00SK20}, the order of the feature requests is not defined in the \ac{ble} specification and can differ for each device. 

After the parameter negotiation, the central initiates the pairing procedure by sending a pairing request to peripheral, which is answered by a pairing response. 
The \ac{ble} protocol distinguishes two pairing methods: \emph{legacy} and \emph{secure} pairing. The difference between the two pairing methods is in the generation of encryption keys. \emph{Legacy} pairing uses priorly exchanged parameters to create a session key, whereas \emph{secure} pairing requires a public/private key exchange to establish a secure encrypted connection. In the remainder of this article, we will only consider \emph{legacy} pairing, since this pairing mode was implemented by all investigated devices.

The \emph{legacy} pairing procedure starts with sending the corresponding request ($\mathsf{legacy\_pairing\_req}$). The \ac{ble} specification \cite{Bluetooth53} refers to the device that sends the pairing request as \emph{initiator} and to the recipient as \emph{responder}. In our case, the central is always the initiator and the peripheral the responder. If the responder accepts to pair, a pairing response ($\mathsf{legacy\_pairing\_rsp}$) is provided. Then both parties generate a  random number and a confirm value, where the calculation of the confirm value takes connection parameters and the random value into account.  First, both parties distribute the confirm value. Next, the initiator sends its generated random value. The responder checks if the confirm and random value of the initiator match. If the values match, the responder returns its random value and the initiator checks whether the values match. Afterwards, the initiator shares its part of the encryption key and the initialization vector by forwarding an encryption request ($\mathsf{encryption\_req}$). The responder first sends the corresponding response ($\mathsf{encryption\_rsp}$) and then a request to start the encryption ($\mathsf{start\_encryption\_req}$). The initiator uses the received key parts and transfers an encrypted response ($\mathsf{start\_encryption\_rsp}$) to the encryption start request. For encryption AES-CCM \cite{DBLP:journals/istr/Murphy99} is used. After the responder receives the encrypted $\mathsf{start\_encryption\_rsp}$, the central and peripheral communication is encrypted. An established communication can be terminated via a termination indication ($\mathsf{termination\_ind}$) and an exchanged encryption key can be renewed by the encryption pause procedure $\mathsf{pause\_encryrption\_req}$.

%% file: mat-framework.tex
\begin{tikzpicture}[thick,font=\scriptsize\sffamily,>=stealth',->]
	\tikzstyle{bigbox} = [draw,font=\footnotesize\sffamily\bfseries,anchor=north,text centered,text depth = 4.9cm]
	\tikzstyle{tinybox} = [draw,font=\scriptsize\sffamily\bfseries ,minimum height = 0.75cm, text width = 1.5cm,text centered]
	
	\node[bigbox,text width =3.4cm] (teacher)  at (0,0){\textcolor{black}{Teacher}};
	\node[bigbox,text width = 1.5cm] (learner)  at (8,0){\textcolor{black}{Learner}};
	
	
	
	\node[tinybox] (ct) at  (-0.6,-1.1) {Conformance Testing};
	\node[tinybox] (mapper) at  (-0.6,-2.8) {Mapper};
	\node[tinybox,fill=black,text=white] (sut) at  (-0.6,-4.6) {System Under Learning};
	
	\node[text width=1.25cm,align=center] (hypothesis) at (2.75,-2.85) {build\\hypothesis};
	\circledarrow{}{hypothesis}{0.69cm};
	
	\node[] (build-hypo-top) at (2.75,-2.2) {};
	\node[] (equivalence-anchor-west) at (4.25,-0.65) {};
	
	\node[] (teacher-top-in) at (1.7,-0.8) {} ;
	\node[] (teacher-top-in-2) at (1.95,-0.8) {} ;
	\node[] (learner-top-out) at (7.25,-0.8) {} ;
	\node[] (ct-top) at (0.15,-0.8) {} ;
	
	\node[] (ct-bottom-out) at (0.15,-1.4) {} ;
	\node[] (ct-bottom-out-2) at (1.95,-1.4) {} ;
	\node[] (teacher-cex-out) at (1.7,-1.4) {} ;
	\node[] (learner-cex-in) at (7.25,-1.4) {} ;
	
	\node[] (perform-in) at (0,-1.2) {} ;
	
	\node[] (query-out) at (7.25,-2.55) {} ;
	\node[] (query-in) at (0.15,-2.55) {} ;
	\node[] (query-in-mq) at (1.5,-2.55) {} ;
	\node[] (query-out-2) at (7.25,-3.1) {} ;
	\node[] (query-in-2) at (0.15,-3.1) {} ;
	\node[] (query-in-2-mq) at (1.5,-3.1) {} ;
	
	\node[] (query-in-short) at (1.5,-2.35) {} ;
	\node[] (query-in-short-2) at (1.5,-2.85) {} ;
	
	\draw[] (learner-top-out) edge (teacher-top-in) node [above left = -0.1cm and 0.25cm, text width = 2.5cm, text centered] {equivalence query};
	\draw[] (teacher-cex-out) edge (learner-cex-in) node [above right= -0.1cm and 2.5cm, text width = 2.5cm, text centered] {yes/counterexample};
	
	\draw[] (query-out) edge (query-in) node [above left = -0.1cm and 0cm, text width = 3.5cm, text centered] {abstract output query \\ + resets};
	\draw[] (query-in-2) edge (query-out-2) node [above right = -0.1cm and 3.75cm, text width = 3cm, text centered] {abstract query output};

	
	
	
	\draw[transform canvas={xshift=-0.6cm}] (mapper) edge (sut) node [below right =0.35cm and -0.05cm, text width = 4.2em, align=left] {output \\ query + resets};
	\draw[transform canvas={xshift=0.6cm}] (sut) edge (mapper) node [above right =0.6cm and -0.05cm,, text width = 4.2em, align=left] {query \\ output};
	
	\draw[transform canvas={xshift=-0.6cm}] (ct) edge (mapper) node [below right =0.3cm and -0.05cm, text width = 4.2em, align=left] {abstract inputs + \\ resets};
	\draw[transform canvas={xshift=0.6cm}] (mapper) edge (ct) node [above right =0.5cm and -0.3cm, text width = 4.2em, text centered] {abstract outputs};
	
	
	\draw[] (teacher-top-in-2) edge (ct-top) node [above left = -0.07cm and -0.075cm,  text width = 1.75cm, text centered] {perform tests};
	\draw[] (ct-bottom-out) edge (ct-bottom-out-2) node [above right = -0.1cm and -0.1cm, text width = 1.75cm, text centered] {passed/failed tests};
	
	\draw[bend left] (build-hypo-top) edge (equivalence-anchor-west);

\end{tikzpicture}

%% file: CC2652R1-model.tex
\begin{tikzpicture}[>=stealth',font=\scriptsize\sffamily,thick,text width=0.4cm] 
	\node[state,initial,initial text=,initial where=above,align=center,font=\footnotesize\bfseries\sffamily,text width=0.25cm,minimum size=1pt] (q0) {q\textsubscript{0}};
	\node[state,below left = 1.1cm and 1.3cm of q0,align=center,font=\footnotesize\bfseries\sffamily,text width=0.25cm,minimum size=1pt] (q1) {q\textsubscript{1}};
	\node[state,below right = 1.1cm and 1.3cm of q0,font=\footnotesize\bfseries\sffamily,text width=0.25cm,minimum size=1pt] (q2) {q\textsubscript{2}};
	\node[state,below = 2.5cm of q0,font=\footnotesize\bfseries\sffamily,text width=0.25cm,minimum size=1pt] (q3) {q\textsubscript{3}};
	
	\path[->] (q0.180) edge[bend right=65,looseness=1.2] node [above left=-0.65cm and 0.25cm,text width=2cm,align=right] {pairing\_req/\\PAIRING\_RSP} (q1.90)
	(q1.45) edge[bend right=0] node [below right=-0.1cm and -0.1cm,text width=2cm] {pairing\_req/\\FAILED} (q0.225)
	(q1.220) edge[bend right=85,looseness=1.45] node [below left=-0.25cm and 0cm,text width= 2cm, align=right] {feature\_rsp/\\LENGTH\_REQ} (q3.200)
	(q3.135) edge[] node [below left=0cm and -0.15cm,text width=1.25cm,align=right] {length\_rsp/\\DATA} (q1.315)
	(q3.0) edge[bend right=65,looseness=1.2] node [below right = -0.5cm and 0.15cm,text width=2.5cm] {pairing\_req/\\FAILED} (q2.270)
	(q2.225) edge[bend right=0] node [above left = 0cm and -0.3cm,text width=2cm,align=right] {pairing\_req/\\PAIRING\_RSP} (q3.45)
	(q0.40) edge[bend left=85,looseness=1.45] node [above left = 0.25cm and -1.75cm,text width=3cm] {feature\_rsp/LENGTH\_REQ} (q2.50)
	(q2.135) edge[bend right=0] node [above right = 0.2cm and -0.5cm,text width=1.5cm] {length\_rsp/\\DATA} (q0.315)
	(q0) edge[loop,in=160,out=105,looseness=4,red] node [above left = -0.1cm and -0.1cm,text width=1.75cm, align=right] {\textcolor{black}{+/+}\\version\_req/\\VERSION\_IND} (q0)
	(q1) edge[loop,in=130,out=200,looseness=4,red] node [left,text width=2cm, align=right] {version\_req/\\VERSION\_IND\\\textcolor{black}{+/+}} (q1)
	(q2) edge[loop,in=325,out=35,looseness=4,red] node [right=-0.05cm,text width=2.45cm] {version\_req/\\VERSION\_IND\\\textcolor{black}{+/+}} (q2)
	(q3) edge[loop,in=240,out=310,looseness=4,red] node [below left= -0.1cm and -0.75cm,text width=2cm] {version\_req/\\VERSION\_IND\\\textcolor{black}{+/+}} (q3)
	;
\end{tikzpicture}

%% file: ble-sequence-diagram.tex
\begin{tikzpicture}[thick,font=\small\sffamily]
	
	\node[draw,text width = 1.75cm,minimum height=0.75cm,text centered,fill=gray!20!white] (central) {\textbf{central} (initiator)};
	\node[draw,text width = 1.75cm,minimum height=0.75cm,text centered,right= 6cm of central,fill=gray!20!white] (peripheral) {\textbf{peripheral} (responder)};
	\node[text width = 1.5cm,minimum height=0.25cm,text centered,below= 10.5cm of central] (central-bot) {};
	\node[text width = 1.5cm,minimum height=0.25cm,text centered,below= 10.5cm of peripheral] (peripheral-bot) {};
	
	\node[text width = 0cm,minimum height=0.75cm,text centered,below right= 0cm and -1cm of peripheral] (ads-start) {};	
	\node[text width = 0cm,minimum height=0.75cm,text centered,below left= 0cm and 2.25cm of peripheral] (ads-end) {};
	
	\node[text width = 0cm,minimum height=0.75cm,text centered,below left= 0.45cm and -1cm of central] (scan-req-start) {};
	\node[text width = 0cm,minimum height=0.75cm,text centered,below right=  0.45cm and -1cm of peripheral] (scan-req-end) {};
	
	\node[text width = 0cm,minimum height=0.75cm,text centered,below = -0.3cm of scan-req-end] (scan-rsp-start) {};
	\node[text width = 0cm,minimum height=0.75cm,text centered,below = -0.3cm of scan-req-start] (scan-rsp-end) {};
	
	\node[text width = 0cm,minimum height=0.75cm,text centered,below = -0.1cm of scan-rsp-end] (con-req-start) {};
	\node[text width = 0cm,minimum height=0.75cm,text centered,below = -0.1cm of scan-rsp-start] (con-req-end) {};
	
	\node[text width = 0cm,minimum height=0.75cm,text centered,below = -0.3cm of con-req-end] (con-rsp-start) {};
	\node[text width = 0cm,minimum height=0.75cm,text centered,below = -0.3cm of con-req-start] (con-rsp-end) {};
	
	\node[text width = 0cm,minimum height=0.75cm,text centered,below = -0.1cm of con-rsp-end] (feature-req-start) {};
	\node[text width = 0cm,minimum height=0.75cm,text centered,below = -0.1cm of con-rsp-start] (feature-req-end) {};
	
	\node[text width = 0cm,minimum height=0.75cm,text centered,below = -0.3cm of feature-req-end] (feature-rsp-start) {};
	\node[text width = 0cm,minimum height=0.75cm,text centered,below = -0.3cm of feature-req-start] (feature-rsp-end) {};
	
	\node[text width = 0cm,minimum height=0.75cm,text centered,below = -0.1cm of feature-rsp-end] (pairing-req-start) {};
	\node[text width = 0cm,minimum height=0.75cm,text centered,below = -0.1cm of feature-rsp-start] (pairing-req-end) {};
	
	\node[text width = 0cm,minimum height=0.75cm,text centered,below = -0.3cm of pairing-req-end] (pairing-rsp-start) {};
	\node[text width = 0cm,minimum height=0.75cm,text centered,below = -0.3cm of pairing-req-start] (pairing-rsp-end) {};

	\node[text width = 0cm,minimum height=0.75cm,text centered,below = -0.1cm of pairing-rsp-end] (confirm-i-start) {};
	\node[text width = 0cm,minimum height=0.75cm,text centered,below = -0.1cm of pairing-rsp-start] (confirm-i-end) {};
	
	\node[text width = 0cm,minimum height=0.75cm,text centered,below = -0.35cm of confirm-i-end] (confirm-r-start) {};
	\node[text width = 0cm,minimum height=0.75cm,text centered,below = -0.35cm of confirm-i-start] (confirm-r-end) {};
	
	\node[text width = 0cm,minimum height=0.75cm,text centered,below = -0.1cm of confirm-r-end] (random-i-start) {};
	\node[text width = 0cm,minimum height=0.75cm,text centered,below = -0.1cm of confirm-r-start] (random-i-end) {};
	
	\node[text width = 0cm,minimum height=0.75cm,text centered,below = -0.35cm of random-i-end] (random-r-start) {};
	\node[text width = 0cm,minimum height=0.75cm,text centered,below = -0.35cm of random-i-start] (random-r-end) {};
	
	\node[text width = 0cm,minimum height=0.75cm,text centered,below = -0.1cm of random-r-end] (encryption-req-start) {};
	\node[text width = 0cm,minimum height=0.75cm,text centered,below = -0.1cm of random-r-start] (encryption-req-end) {};
	
	\node[text width = 0cm,minimum height=0.75cm,text centered,below = -0.35cm of encryption-req-end] (encryption-rsp-start) {};
	\node[text width = 0cm,minimum height=0.75cm,text centered,below = -0.35cm of encryption-req-start] (encryption-rsp-end) {};
	
	\node[text width = 0cm,minimum height=0.75cm,text centered,below = -0.1cm of encryption-rsp-start] (start-encryption-req-start) {};
	\node[text width = 0cm,minimum height=0.75cm,text centered,below = -0.1cm of encryption-rsp-end] (start-encryption-req-end) {};
	
	\node[text width = 0cm,minimum height=0.75cm,text centered,below = -0.35cm of start-encryption-req-end] (start-encryption-rsp-start) {};
	\node[text width = 0cm,minimum height=0.75cm,text centered,below = -0.35cm of start-encryption-req-start] (start-encryption-rsp-end) {};
	
	\node[text width = 0cm,minimum height=0.75cm,text centered,below = -0.1cm of start-encryption-rsp-start] (encryption-info-start) {};
	\node[text width = 0cm,minimum height=0.75cm,text centered,below =  -0.1cm of start-encryption-rsp-end] (encryption-info-end) {};

	\draw
	(central.south) edge[->,>=latex] (central-bot.north)
	(peripheral.south) edge[->,>=latex] (peripheral-bot.north)
	
	(ads-start.west) edge[->,>=latex] (ads-end.east) node [above left = 0cm and 0.5cm, text width = 2cm] {advertisements}
	
	(scan-req-start.east) edge[->,>=latex] (scan-req-end.west) node [above right = -0.1cm and 3.25cm, text width = 2cm] {scan\_req}
	
	(scan-rsp-start.west) edge[->,>=latex,dashed] (scan-rsp-end.east) node [above left = -0.1cm and 2.5cm, text width = 2cm] {scan\_rsp}
	
	(con-req-start.east) edge[->,>=latex] (con-req-end.west) node [above right = -0.1cm and 2.75cm, text width = 2cm] {connection\_req}
	
	(con-rsp-start.west) edge[->,>=latex,dashed] (con-rsp-end.east) node [above left = -0.1cm and 3cm, text width = 2cm] {connection\_rsp}
	
	(feature-req-start.east) edge[<->,>=latex] (feature-req-end.west) node [above right = -0.1cm and 1.25cm, text width = 7cm] {\{length, feature, version, MTU\}\_req}
	
	(feature-rsp-start.west) edge[<->,>=latex,dashed] (feature-rsp-end.east) node [above left = -0.1cm and -0.5cm, text width = 7cm] {\{length, feature, version, MTU\}\_rsp}
	
	(pairing-req-start.east) edge[->,>=latex] (pairing-req-end.west) node [above right = -0.1cm and 2.5cm, text width = 7cm] {legacy pairing\_req}
	
	(pairing-rsp-start.west) edge[->,>=latex,dashed] (pairing-rsp-end.east) node [above left = -0.1cm and -0.75cm, text width = 6cm] {legacy pairing\_rsp}
	
	(confirm-i-start.east) edge[->,>=latex] (confirm-i-end.west) node [above right = -0.1cm and 3.25cm, text width = 2cm] {confirm\textsubscript{i}}
	
	(confirm-r-start.west) edge[->,>=latex,dashed] (confirm-r-end.east) node [above left = -0.1cm and 2.5cm, text width = 2cm] {confirm\textsubscript{r}}
	
	(random-i-start.east) edge[->,>=latex] (random-i-end.west) node [above right = -0.1cm and 3.25cm, text width = 2cm] {random\textsubscript{i}}
	
	(random-r-start.west) edge[->,>=latex,dashed] (random-r-end.east) node [above left = -0.1cm and 2.5cm, text width = 2cm] {random\textsubscript{r}}
	
	(encryption-req-start.east) edge[->,>=latex] (encryption-req-end.west) node [above right = -0.1cm and 2.75cm, text width = 2cm] {encryption\_req}
	
	(encryption-rsp-start.west) edge[->,>=latex,dashed] (encryption-rsp-end.east) node [above left = -0.1cm and 3cm, text width = 2cm] {encryption\_rsp}
	
	(start-encryption-req-start.west) edge[->,>=latex,dashed] (start-encryption-req-end.east) node [above left = -0.1cm and 3.25cm, text width = 2cm] {start\_encryption\_req}
	
	(start-encryption-rsp-start.east) edge[->,>=latex] (start-encryption-rsp-end.west) node [above right = -0.1cm and 2.4cm, text width = 2cm] {$\langle$start\_encryption\_rsp$\rangle$}
	
	(encryption-info-start.east) edge[<->,>=latex] (encryption-info-end.west) node [above right = -0.1cm and 0.5cm, text width = 8cm] {$\langle$encryption\_info, signing\_info, master\_identification$\rangle$}
	;

	\node[draw,dashed,text width = 1.5cm,minimum height=0.75cm,text centered,rounded corners,below left = 0.85cm and -0.5cm of central] (scanning) {scanning};
	\node[draw,dashed,text width = 1.5cm,minimum height=0.75cm,text centered,rounded corners,below = 0.5cm of scanning] (initiating) {initiating};
	\node[draw,dashed,text width = 1.5cm,minimum height=0.75cm,text centered,rounded corners,below = 0.5cm of initiating] (connectionM) {connection};
	\node[draw,dashed,text width = 1.5cm,minimum height=0.75cm,text centered,rounded corners,below = 0.55cm of connectionM] (randomI) {random\textsubscript{i}, confirm\textsubscript{i} generation};
	\node[draw,dashed,text width = 1.5cm,minimum height=0.75cm,text centered,rounded corners,below = 0.75cm of randomI] (check-confirm-i) {check confirm};
	\node[draw,dashed,text width = 1.5cm,minimum height=0.75cm,text centered,rounded corners,below = 0.3cm of check-confirm-i] (sk-i) {session key generation};
	\node[draw,dashed,text width = 1.5cm,minimum height=0.75cm,text centered,rounded corners,below = 0.3cm of sk-i] (encryption-i) {encryption enabled};
	
		\draw
	(scanning.south) edge[->,>=stealth'] (initiating.north)
	(initiating.south) edge[->,>=stealth'] (connectionM.north)
	(connectionM.south) edge[->,>=stealth'] (randomI.north)
	(randomI.south) edge[->,>=stealth'] (check-confirm-i.north)
	(check-confirm-i.south) edge[->,>=stealth'] (sk-i.north)
	(sk-i.south) edge[->,>=stealth'] (encryption-i.north)
	;
	
	\node[draw,dashed,text width = 1.5cm,minimum height=0.75cm,text centered,rounded corners,below right = 0.15cm and -0.5cm of peripheral] (advertising) {advertising};
	\node[draw,dashed,text width = 1.5cm,minimum height=0.75cm,text centered,rounded corners,right = 9.025cm of connectionM] (connectionP) {connection};
	\node[draw,dashed,text width = 1.5cm,minimum height=0.75cm,text centered,rounded corners,below = 0.55cm of connectionP] (randomR) {random\textsubscript{r}, confirm\textsubscript{r} generation};
	\node[draw,dashed,text width = 1.5cm,minimum height=0.75cm,text centered,rounded corners,below = 0.3cm of randomR] (check-confirm-r) {check confirm};
	\node[draw,dashed,text width = 1.5cm,minimum height=0.75cm,text centered,rounded corners,right = 9.025cm of sk-i] (sk-r) {session key generation};
	\node[draw,dashed,text width = 1.5cm,minimum height=0.75cm,text centered,rounded corners,below = 0.3cm of sk-r] (encryption-r) {encryption enabled};
	
	\draw
	(advertising.south) edge[->,>=stealth'] (connectionP.north)
	(connectionP.south) edge[->,>=stealth'] (randomR.north)
	(randomR.south) edge[->,>=stealth'] (check-confirm-r.north)
	(check-confirm-r.south) edge[->,>=stealth'] (sk-r.north)
	(sk-r.south) edge[->,>=stealth'] (encryption-r.north)
	;
	
\end{tikzpicture}

%
%
%
%
%
%

%% file: architecture.tex



Our objective is to learn the behavioral model of the \ac{ble} protocol implemented by the peripheral device. The learning setup is based on active automata learning. \new{Following related protocol state fuzzing techniques \cite{DBLP:conf/uss/RuiterP15,DBLP:conf/cav/Fiterau-Brostean16,DBLP:conf/spin/Fiterau-Brostean17,DBLP:conf/icst/TapplerAB17,DBLP:conf/esorics/StoneCR18,DBLP:conf/uss/Fiterau-Brostean20}, we assume that unusual input sequences that are executed during active learning test the robustness of the \ac{sul}.} Additionally, we aim to reveal characteristic behavior that enables fingerprinting of the peripheral. According to Section~\ref{sec:ble}, we can model the \ac{ble} protocol as a reactive system.

\new{Our objective is to learn the behavior during the connection and pairing procedure as depicted in Figure~\ref{fig:ble-communication}. However, the first experiments indicate that especially the pairing procedure leads more frequently to non-deterministic behavior which hampers the learnability of \ac{ble} devices. Therefore, we separated the learning of the connection and pairing procedure. Our first learning setup considers inputs that are required to establish a connection until a pairing request initiates the pairing procedure. Related to Figure~\ref{fig:ble-communication}, this connection procedure includes the first four request/response steps. The second learning setup considers only the inputs required to establish an encrypted communication, i.e., the request/response steps of the pairing procedure, including the pairing request. Hence, both models contain the pairing request.}

\begin{figure}[t]
	\centering
	\input{interface}
	\caption{\new{We extended the learning architecture of Tappler et al.~\cite{DBLP:conf/icst/TapplerAB17} with a learning interface that enables the robust learning of an abstracted model from a \ac{ble} device.}}
	\label{fig:interface}
\end{figure}
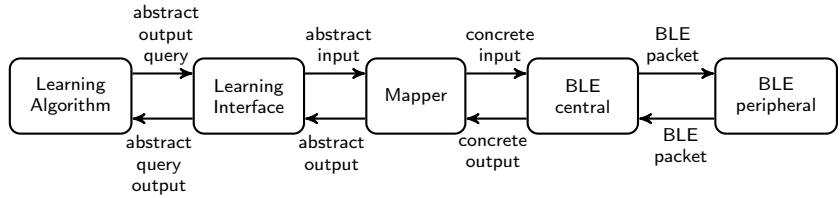

Even though the considered inputs are different, we apply the same learning architecture for both learning setups. Figure~\ref{fig:interface} depicts our learning architecture which is based on the architecture for network protocols proposed by Tappler et al.~\cite{DBLP:conf/icst/TapplerAB17}. 
\new{We extended their proposed learning architecture with an additional learning interface. This learning interface should enable the robust learning of a behavioral model. This additional layer ensures an reliable reset and query execution on the \ac{sul}.}  Figure~\ref{fig:interface} includes the five components of the learning interface: learning algorithm (Section~\ref{sec:learning-algorithm}), learning interface (Section~\ref{sec:learning-interface}), mapper (Section~\ref{sec:mapper}), \ac{ble} central (Section~\ref{sec:ble-central}), and \ac{ble} peripheral (Section~\ref{sec:ble-peripheral}). In the following sections, we describe each component of the applied learning architecture.

\subsection{Learning Algorithm}\label{sec:learning-algorithm}

The applied \emph{learning algorithm} is an improved variant of the $L^*$ algorithm for Mealy machines \cite{DBLP:conf/hldvt/MargariaNRS04,DBLP:phd/de/Niese2003,DBLP:conf/fm/ShahbazG09}. Since $L^*$ is based on an exhaustive input exploration in each state, we assume that it is beneficial for performing a behavioral analysis and fingerprinting. Rivest and Schapire \cite{DBLP:journals/iandc/RivestS93} proposed the improved $L^*$ version that contains advanced counterexample processing.
\new{This advanced counterexample processing reduces the number of required output queries in case a counterexample to the conformance between the \ac{sul} and provided hypothesis is found. Especially for long counterexamples, this improved $L^*$ version decreases the number of required queries.}

Since Python enables the usage of convenient libraries for the composition of \ac{ble} packets, we aim at a consistent learning framework integration. At present, \textsc{AALpy} \cite{aalpy} is a novel active learning library written in Python. \textsc{AALpy} implements state-of-the-art learning algorithms and conformance testing techniques, including the improved $L^*$ variant that is considered here. Since the framework implements equivalence queries via conformance testing, we assume that the conformance relation defined in Equation~\ref{eq:conformance-relation} holds. To create a sufficient test suite, we combine random testing with state coverage. \new{For this, the generated set of input sequences accesses each state in the learned hypothesis. In each state, we then execute a random set of inputs.}  
The applied test-case generation technique generates for each state in the hypothesis $n_\mathrm{test}$ input traces. The generated input traces of length $n_\mathrm{len}$ comprise the input prefixes to the currently considered state concatenated with a random input sequence. \new{Since the final number of states of the minimal Mealy machine that represents the \ac{sul} is unknown, the parameters $n_\mathrm{test}$ and $n_\mathrm{len}$ can only be approximated. The approximation is a trade-off between sufficient conformance testing to find a counterexample and an efficient runtime of the learning algorithm.}

\subsection{Learning Interface}\label{sec:learning-interface}

The applied $L^*$-based learning algorithm requires the system to be resettable and to behave deterministically. These requirements hamper the straightforward application of learning physical devices via a wireless network. \new{To overcome these issues, we introduce an additional layer that ensures that the \ac{sul} is reliably reset and that the observed non-deterministic behavior is resolved.} 

\subsubsection{Guarantee Reliable Resets}

The learning algorithm expects that every performed query is executed from an initial state. Hence, we require that the \ac{sul} can be reset to an initial state. \new{Lee and Yannakakis~\cite{DBLP:journals/pieee/LeeY96} defined a reset to be \emph{reliable} if this reset action transfers the system to the initial state independent from the current state. For \ac{ble} device a manual hard reset could is considered to be reliable.} Since a hard reset after each query would make active learning a tedious process, we assume that the central can reset the device via \ac{ble} messages. For this, we assume that a $\mathsf{scan\_req}$ or a $\mathsf{termination\_req}$ resets the peripheral in any state to the advertising state as shown in Figure~\ref{fig:ble-communication}. \new{Depending on which part of the protocol we aim to learn, i.e., the connection or the pairing procedure, we consider different initial states. The initial state for the connection procedure is the advertising state. For the pairing procedure, we assume that the connection is established and all required parameters are negotiated such that the pairing procedure can be initiated.}

The learning library \textsc{AALpy} can perform resetting actions before and after the output query execution. \new{Like in other automata learning libraries \cite{DBLP:conf/cav/IsbernerHS15}}, we denote the method that is called \emph{before} executing the output query as \texttt{pre} and the method \emph{after} the output query as \texttt{post}. \new{In any case, we want to terminate a possibly established connection after the execution of an output query.} For this, we perform a termination request in the \texttt{post} method. \new{The termination request is specific \ac{ble} packet that indicates the device that sends this indication wants to terminate the connection.} To ensure a proper reset before executing the output query, a scan request is performed in the \texttt{pre} method which checks if the device distributes advertisements. 

For learning the pairing procedure, we extend the \texttt{pre} and \texttt{post} method. In the \texttt{pre} method, we establish a valid connection, which includes the first three steps of Figure~\ref{fig:ble-communication}. After this procedure, the peripheral should be ready to accept a pairing request from the central. In the \texttt{post} method, we initiate the pause encryption procedure if encryption is enabled. This procedure indicates that the encryption key shall be changed. Afterward, we terminate the connection as described before.

\new{We assume that this reset procedure reliably resets the \ac{sul} to the assumed initial state if the peripheral responds to our resetting \ac{ble} messages. We if do not receive a response, we repeat this resetting procedure $n_\mathrm{error}$ times. For example, we repeat the \texttt{pre} method as long as we found advertisements sent by the peripheral. After $n_\mathrm{error}$ repetitions, we abort the learning procedure. }

\new{Another problem that hampers a reset is that some devices stop sending advertisements. This could be the case if we send a large number of unexpected \ac{ble} packets to the device, even if we are not connected. In active automata learning this might be the case, e.g., if we test conformance. For this, we use the resetting procedure also to establish and terminate a valid connection before we execute an output query. This should avoid the peripheral from running into a timeout and stops sending advertisements.} 


\subsubsection{Handling Non-determinism}

Our learning interface also has to deal with non-deterministic behavior. \new{In general, we assume that the \ac{ble} devices behave deterministically. However, the influence of environmental conditions on learning a wireless communication protocol can introduce non-determinism. In learning a wireless communication protocol, we might experience lost packets or delayed responses. Packet loss is critical for packets that are necessary to establish a connection. For example, the loss of a connection request packet prohibits the establishment of a connection between the two \ac{ble} devices. As a result, all responses to further requests are answered differently than for a valid connection.}  
\new{The same problem arises for \ac{ble} packets that arrive delayed. For example, consider the following output query.}
\[
\mathsf{scan\_req} \cdot \mathsf{connection\_req} \cdot \mathsf{feature\_rsp}  \cdot \mathsf{pairing\_req}
\]
\new{The expected query output would be the following.}
\[
\mathsf{ADV} \cdot \mathsf{FEATURE\_REQ} \cdot \mathsf{DATA}  \cdot \mathsf{PAIRING\_RSP}
\]
\new{However, if the peripheral device receives the feature response delayed or if the response got lost, the peripheral might reject the pairing request. Hence, the received query output would look different. In this case, a possible example would be the following output sequence.}
\[
\mathsf{ADV} \cdot \mathsf{FEATURE\_REQ} \cdot \mathsf{DATA}  \cdot \mathsf{FAILED}
\]


\new{To deal with non-deterministic behavior that occurs during the execution of output queries, we repeat output queries. However, repeating every output query several times would make learning very inefficient. For this, we introduce an enhanced strategy that aims at the saving of query. Under the assumption that packet loss or delayed messages occur rarely, we repeat queries only if we observe non-deterministic behavior.} For this, our learning interface utilizes the caching strategy of the used learning library \textsc{AALpy}. \textsc{AALpy} provides a tree structure that collects the performed inputs and the corresponding observed outputs. Every observed output on the \ac{sul} is checked against the stored output in the cache. We start to collect possible outputs for a node in the tree only in the case that the outputs do not match. After the collection of $n_\mathrm{cache}$ outputs for that node, we select the most frequently observed output. If the output changes, the cache gets updated. \new{Note that the update might violate the consistency of the data structure for learning. However, the performed counterexample processing proposed by Rivest and Schapire~\cite{DBLP:journals/iandc/RivestS93} during the conformance test takes care of any inconsistencies.} The majority-based update is only done once. Afterward, if we observe non-determinism, we simply repeat the output query. Again, we define an upper limit for a repeating non-deterministic behavior by a maximum of $n_\mathrm{nondet}$ query executions.

\subsection{Mapper}\label{sec:mapper}


The \emph{mapper} component serves as an abstraction mechanism, since considering all possible \ac{ble} packets for learning would not be feasible.  Therefore, we use a generic input and output alphabet to learn a behavioral model on a more abstract level. Following Figure~\ref{fig:interface}, the learning algorithm generates output queries that comprise abstract input sequences. \new{The learning interface receives these abstract input sequences and forwards the single abstract inputs to the mapper. The mapper then translates them to concrete inputs that can be executed by the central. After the central received a concrete input action, the central returns the corresponding concrete output. This concrete output is then taken by the mapper and translated to a more abstract output that is used by the learning interface to perform the corresponding actions for robust learning.} The processed abstract output sequence is then used by the learning algorithm to construct and test the hypothesis.

The abstracted input alphabet for learning the connection procedure is defined by $I_C^\mathcal{A} = \{ \mathsf{scan\_req}, \mathsf{connection\_req}, \mathsf{length\_req}, \mathsf{length\_rsp}, \mathsf{feature\_req},$ $\mathsf{feature\_rsp}, \mathsf{version\_req}, \mathsf{mtu\_req}, \mathsf{legacy\_pairing\_req} \}$ and for the pairing procedure $I_P^\mathcal{A} = \{\mathsf{legacy\_pairing\_req}, \mathsf{confirm}, \mathsf{random}, \mathsf{encryption\_req},\hspace*{2cm}$ $\mathsf{start\_encryption\_rsp} \}$. Considering the input/output definition of reactive systems, it may be unusual to include responses in the input alphabet. For our setup, we included the feature and length response as inputs. In Section~\ref{sec:ble}, we explained that after the connection request of the central, also the peripheral might send control packets or feature requests. To explore more behavior of the peripheral, we have to reply to received requests from the peripheral. In a learning setup, the inputs $\mathsf{feature\_rsp}$, $\mathsf{length\_rsp}$, and $\mathsf{start\_encryption\_rsp}$  are responses from the central that we consider as additional inputs.

The abstract inputs of $I_C^\mathcal{A}$ and $I_P^\mathcal{A}$ are then translated to concrete \ac{ble} packets that can be sent by the central to the peripheral. For example, the abstract input $\mathsf{length\_req}$ is translated to a \ac{ble} control packet including a corresponding valid command of the \ac{ble} protocol stack. For the construction of the \ac{ble} packets, we use the Python library \textsc{Scapy} \cite{scapy}. In \textsc{Scapy} syntax, the \ac{ble} packet for the $\mathsf{length\_req}$  can be defined as 

\[
	\mathsf{BTLE} / \mathsf{BTLE\_DATA} / \mathsf{BTLE\_CTRL} / \mathsf{LL\_LENGTH\_REQ}(\mathit{max\_tx\_bytes}, \ldots),
\]

where $\mathit{max\_tx\_bytes}$ is a field that is concretized by the mapper component. \new{To concretize fields, we mainly select values from a set of preset values defined by \textsc{Scapy}. These preset values conform to standard values that enable the establishment of a connection.} 

\new{For the translation of outputs, the mapper receives a list of concrete \ac{ble} packets from the central device. The central provides a list of packets since the peripheral device answers with multiple \ac{ble} packets to a single \ac{ble} request. The set of received packets is parsed using the \textsc{Scapy} library. In the following example, we receive for the sent $\mathsf{feature\_rsp}$ a list of different \ac{ble} packets.}

\hspace*{-0.8cm}
\new{
{
\footnotesize
$\begin{array}{lrl}
\mathit{req} & = & \mathsf{BTLE} / \mathsf{BTLE\_DATA} / \mathsf{BTLE\_CTRL} / \mathsf{LL\_FEATURE\_RSP} \\
\mathit{rsp} & = & \{ \mathsf{BTLE} / \mathsf{BTLE\_DATA},\\
& & \mathsf{BTLE} / \mathsf{BTLE\_DATA} / \mathsf{BTLE\_CTRL} / \mathsf{LL\_LENGTH\_REQ},\\
& & \mathsf{BTLE} / \mathsf{BTLE\_DATA},\\
& & \mathsf{BTLE} / \mathsf{BTLE\_DATA},\\
& & \mathsf{BTLE} / \mathsf{BTLE\_DATA} / \mathsf{L2CAP\_Hdr} / \mathsf{ATT\_Hdr} / \mathsf{ATT\_Exchange\_MTU\_Request},\\ 
& & \mathsf{BTLE} / \mathsf{BTLE\_DATA},\\
& & \mathsf{BTLE} / \mathsf{BTLE\_DATA} \}
\end{array}$
}}

\new{For simplicity, the example hides concrete field values and illustrates the received \ac{ble} packets conforming to the \textsc{Scapy} syntax. This list of output packets will be merged into a single output by concatenating the packets in alphabetical order to one output string. This creates deterministic behavior, even though packets might be received in a different order. For the example above, the output would be $\mathsf{ATT\_Exchange\_MTU\_Request,ATT\_Hdr,BTLE,BTLE\_CTRL,}\hspace*{1cm}$ $\mathsf{BTLE\_DATA,L2CAP\_Hdr,LL\_LENGTH\_REQ}$. For the abstraction of the \ac{ble} packets, we use the naming provided by \textsc{Scapy}. One exception applies to the response on $\mathsf{scan\_req}$, where two possible valid responses are mapped to one scan response ($\mathsf{ADV}$). If the central device returns an empty list of \ac{ble} packets, the mapper returns the empty output which is denoted by the string $\mathsf{EMPTY}$.}




\new{In the pairing procedure, the central and the peripheral device exchange key information over a multistage response/request dialog. Since the concrete values of the packets in this key-exchange procedure depend on the previously exchanged packets, we cannot use preset values from \textsc{Scapy} for the concretization. Also randomly guessing the concrete values for the key exchange would not be feasible to successfully establish an encrypted connection. Otherwise, the key-exchange procedure would be insecure, since the keys do not depend on randomness or could be brute-forced.} Due to this key exchange, we require the mapper to be stateful. For this, the mapper stores and collects messages that are later required to establish an encrypted communication, e.g., received parts of the key information. Furthermore, the mapper memorizes if the encryption is enabled. In the case of encrypted communication, the mapper encrypts and decrypts transmitted messages.

\subsection{BLE Central}\label{sec:ble-central}
The \emph{\ac{ble} central} component comprises the adapter implementation and the physical central device. We use the Nordic nRF52840 USB dongle and the Nordic nRF52840 Development Kit as central. Our learning setup requires sending \ac{ble} packets stepwise to the peripheral device. For this, our implementation follows the setup proposed by Garbelini et al.~\cite{DBLP:conf/usenix/Garbelini00SK20}. We use their provided firmware for the Nordic nRF52840 devices and adapted their driver implementation to perform single steps of the \ac{ble} protocol. 

\new{The central device receives from the mapper a concrete \ac{ble} packet, which is then transmitted to the peripheral device. Then the central device check for responses if the peripheral responds to the transmitted packet. As mentioned in the previous section, the peripheral can respond with several packets. For this, the central listens $n^\mathrm{rsp}_\mathrm{min}$ times for any responses. If after $n^\mathrm{rsp}_\mathrm{min}$ responses no convincing response has been returned, we continue listening for responses. We define a response as \emph{convincing} if the received packet contains more than an empty \ac{ble} data packet, i.e., $\mathsf{BTLE} / \mathsf{BTLE\_DATA}$.  However, the maximum number of listening attempts is limited by $n^\mathrm{rsp}_\mathrm{max}$. The selection of the parameters $n^\mathrm{rsp}_\mathrm{min}$ and $n^\mathrm{rsp}_\mathrm{max}$ depends on the environmental conditions in which the experiment is executed. For example, we need to consider the response time and distance of the \ac{sul}.}

\subsection{BLE Peripheral}\label{sec:ble-peripheral}
The \emph{\ac{ble} peripheral} represents the black-box device that we want to learn, i.e., the \ac{sul}. We assume that the peripheral advertises and only interacts with our central device. For learning, we require that the peripheral is resettable and that the reset can be initiated by the central. After a reset, the peripheral should be again in the advertising state.

%% file: interface.tex
\begin{tikzpicture}[thick,font=\scriptsize\sffamily]
	
	\node[draw,rounded corners,text width = 1.4cm,minimum height=1cm,text centered] (algo) {Learning Algorithm};
	
	\node[draw,rounded corners,text width = 1.25cm,minimum height=1cm,text centered, right=0.8cm of algo] (interface) {Learning Interface};

	\node[draw,rounded corners,text width = 1.1cm,minimum height=1cm,text centered, right=0.8cm of interface] (mapper) {Mapper};
	
	\node[draw,rounded corners,text width = 1.25cm,minimum height=1cm,text centered, right=0.8cm of mapper] (client) {BLE central};
	
	\node[draw,rounded corners,text width = 1.4cm,minimum height=1cm,text centered, right=1cm of client] (broker) {BLE peripheral};
	
	\path[draw]
	
	(algo.0) edge[->,>=stealth',transform canvas={yshift=0.3cm}] (interface.180) node [above right = 0.3cm and -0.3cm, text width = 1.25cm, text centered] {abstract\\ output\\ query}
	(interface.180) edge[->,>=stealth',transform canvas={yshift=-0.3cm}] (algo.0) node [below left = 0.4cm and -0.3cm, text width = 1.25cm, text centered] {abstract\\ query\\ output}
	(interface.0) edge[->,>=stealth',transform canvas={yshift=0.3cm}] (mapper.180) node [above right = 0.3cm and -0.3cm, text width = 1.25cm, text centered] {abstract\\ input}
	(mapper.180) edge[->,>=stealth',transform canvas={yshift=-0.3cm}] (interface.0) node [below left = 0.4cm and -0.3cm, text width = 1.25cm, text centered] {abstract\\ output}
	(mapper.0) edge[->,>=stealth',transform canvas={yshift=0.3cm}] (client.180) node [above right = 0.3cm and -0.3cm, text width = 1.25cm, text centered] {concrete\\ input}
	(client.180) edge[->,>=stealth',transform canvas={yshift=-0.3cm}] (mapper.0) node [below left = 0.4cm and -0.3cm, text width = 1.25cm, text centered] {concrete\\ output}
	(client.0) edge[->,>=stealth',transform canvas={yshift=0.3cm}] (broker.180) node [above right = 0.3cm and -0.3cm, text width = 1.25cm, text centered] {BLE\\ packet}
	(broker.180) edge[->,>=stealth',transform canvas={yshift=-0.3cm}] (client.0) node [below left = 0.3cm and -0.3cm, text width = 1.25cm, text centered] {BLE\\ packet}
	;
\end{tikzpicture}

%% file: case-study.tex
We evaluated the proposed automata learning setup for the \ac{ble} protocol in a case study consisting of six different \ac{ble} devices.
The learning framework is available \new{online}~\cite{bleLearning}. 
The repository contains the source code for the \ac{ble} learning framework, the firmware for the Nordic nRF52840 Dongle and Nordic nRF52840 Development Kit, the learned automata, and the learning results. 

\subsection{\ac{ble} Devices}

Table~\ref{tab:devices} lists the six investigated \ac{ble} devices. In the remainder of this section, we refer to the \ac{ble} devices by their \ac{soc} identifiers. For the case study, we considered devices from different manufacturers. Some of the devices were already included in the case study of Garbelini et al.~\cite{DBLP:conf/usenix/Garbelini00SK20}. We extended the collection with well-known boards, e.g., the Raspberry Pi 4. Since one of our objectives is to identify the \ac{soc} based on the observed behavior, we included different \acp{soc} from one manufacturer. All evaluated \acp{soc} support the Bluetooth v5.0 standard \cite{Bluetooth53}.  To enable \ac{ble} communication, we deployed and ran an exemplary \ac{ble} application on the \ac{soc}. The considered \ac{ble} applications were either already installed by the semiconductor manufacturer or taken from examples in the semiconductor's specific software development kits.

\begin{table}[t]
	\centering
	\caption{Evaluated \ac{ble} devices}\label{tab:devices}
	\resizebox{\textwidth}{!}{
		\input{devices}}
\end{table}

\subsection{\ac{ble} Learning}

Our learning framework is built upon Python 3.9.0. For our learning setup, we used the Python learning library \textsc{AALpy} \cite{aalpy} (version 1.0.1). For the composition of the \ac{ble} packets, we used a modified version of the Python library \textsc{Scapy} \cite{scapy} (version 2.4.4). The used modifications are available starting from \textsc{Scapy} v2.4.5. As \ac{ble} central device, we used the Nordic nRF52840 Dongle and the Nordic nRF52840 Development Kit. The deployed firmware for the USB dongle was taken from the \textsc{SweynTooth} repository \cite{sweyntooth}. 


As explained in the previous section, we required a special learning interface to learn the communication protocol implemented on a physical device. For this, we extended some components of the \textsc{AALpy} framework to enable robust automata learning. Our learning interface modified the implementation of the conformance testing technique and the used caching mechanism. These modifications of our framework handled connection errors and non-deterministic outputs according to our explanation in Section~\ref{sec:learning-setup}. To enable an efficient but also robust learning environment, we set the maximum number of consecutive connection errors to $n_\mathrm{error} = 20$, the size of the non-deterministic cache to $n_\mathrm{cache} = 20$, and the number of consecutive non-deterministic output queries to $n_\mathrm{nondet} = 20$. \new{These numbers were high enough to recover from faults but low enough to detect early that a device stopped responding.}

For conformance testing, we copied the class \texttt{StatePrefixEqOracle} from \textsc{AALpy} and added our error-handling behavior. The number of performed queries per state is set to $n_\mathrm{test} = 10$ and the number of performed inputs per query is set to $n_\mathrm{len} = 10$. \new{These numbers were set according to the abstract input alphabet size, which included nine different inputs.}  We stress that the primary focus of this article was an initial exploration of the state space and to fingerprint the investigated \ac{ble} \acp{soc}. Therefore, it was sufficient to perform a lower number of conformance tests. However, we recommend increasing the number of conformance tests if a more accurate statement about the conformance of the model to the \ac{sul} is required. 

In Section~\ref{sec:learning-setup}, we explained that a sent \ac{ble} message  leads to multiple responses. These responses could be distributed over several \ac{ble} packets. Hence, our central listened for a minimum number of responses $n^\mathrm{rsp}_\mathrm{min}$ but stopped listening after $n^\mathrm{rsp}_\mathrm{max}$ attempts. For our learning setup, we set for five out of six \acp{soc} $n^\mathrm{rsp}_\mathrm{min} = 10$ and $n^\mathrm{rsp}_\mathrm{max} = 20$. This setup enabled robust and fast learning for five \acp{soc}, \new{since none of the devices responded with more than ten different outputs, but was able to send a meaningful response within twenty listening attempts}. For the sixth device, the nRF52832, we used $n^\mathrm{rsp}_\mathrm{min} = 20$ and $n^\mathrm{rsp}_\mathrm{max} = 30$ since our experiments show that this device requires more time to respond. Furthermore, we applied a different parameter setup for the scan request and termination indication that enables a fast, but decent reset. For this, we set  $n^\mathrm{rsp}_\mathrm{min} = 5$ and $n^\mathrm{rsp}_\mathrm{max} = 50$. For the termination indication, we set $n^\mathrm{rsp}_\mathrm{min} = n^\mathrm{rsp}_\mathrm{max} = 1$. \new{Note that the termination indication was not part of the abstract input alphabet. Hence, the purpose was not to capture the output behavior on this request but simply to reset the connection. This justified the low number of listening attempts for the output.}

\subsubsection{Connection Procedure Evaluation}

All experiments for learning the connection procedure were performed  on an Apple MacBook Pro 2019 with an Intel Quad-Core i5 operating at \unit[2.4]{GHz} and with \unit[8]{GB} RAM, running macOS Catalina (Version 10.15.7).

\begin{table}[t]
	\centering
	\caption{Learning results of five out of six evaluated \ac{ble} \acp{soc}. The $\dagger$-symbol indicates that device was reset to the state where a connection request was already performed. Consequently, no connections errors occur during learning for these devices.}\label{tab:learning-results}
	\resizebox{\textwidth}{!}{
	\input{learning-results}}
\end{table}

\begin{figure}[t]
	\input{automaton-cc2650.tex}
	\caption{Simplified learned model of the CC2650. Inputs are lowercased and outputs are capitalized. For a clear presentation, outputs are abbreviated and we highlight the behavior on some selected input/output labels. Other inputs and outputs are summarized by the $\mathsf{+}$-symbol. The complete model is available online~\cite{bleLearning}.}
	\label{fig:learned-model-CC2650}
\end{figure}
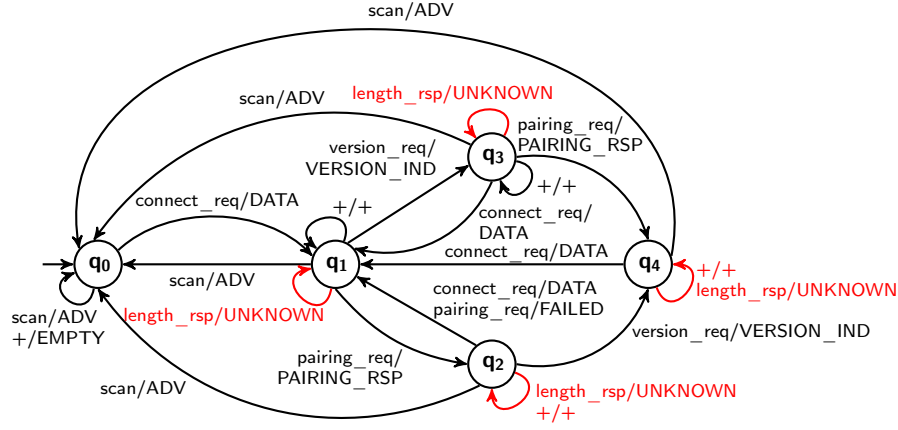
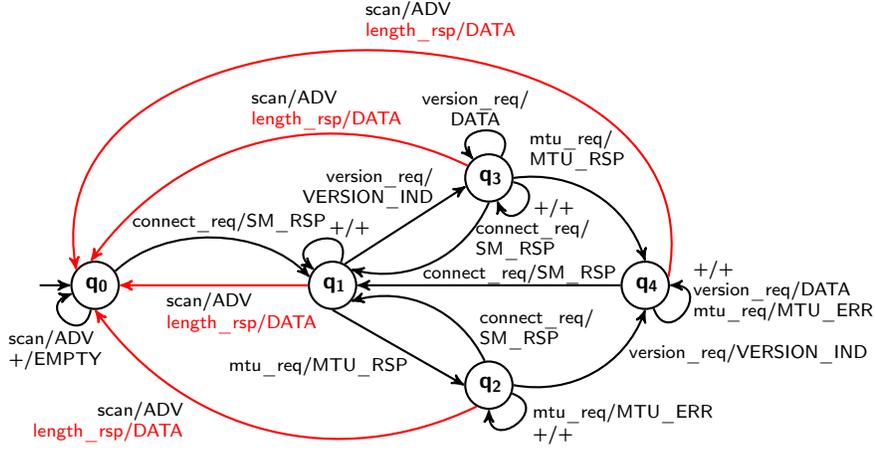
\begin{figure}[t]
	\input{automaton-nrf52832.tex}
	\caption{Simplified learned model of the nRF52832. Labels are abbreviated and summarized. The complete model is available online~\cite{bleLearning}. \new{Unlike the model of CC2650, presented in Figure~\ref{fig:learned-model-CC2650}, the nRF52832 reset the connection after an unexpected length response.}}
	\label{fig:learned-model-nRF52832}
\end{figure}

Table~\ref{tab:learning-results} shows the learning results for five out of the six investigated \acp{soc}. For two devices, CC2652R1 and CYW43455, we excluded the scan and connection request from the input alphabet. \new{For both devices, we observed that a scan request did not always trigger the expected reset of the connection. Therefore, we needed to check if a connection could be established.} For this, we learned the behavior after the execution of a connection request. Table \ref{tab:learning-results} does not include the results of CC2640R2, since we were not able to learn a deterministic model of CC2640R2 using the defined input alphabet. We discuss possible reasons for the non-deterministic behavior later. For all other \acp{soc}, we learned a deterministic Mealy machine using the complete input alphabet.

\new{We needed only one learning round for each \ac{sul}}, i.e., we did not find a counterexample to conformance between the initially created hypothesis and the \ac{sul}. The learned behavioral models range from a simpler structure with only three states (CYBLE-416045-02) to more complex behavior that can be described by 16 states (CYW43455). 

The learning of the largest model regarding the number of states (CYW43455) took a bit more than one hour, whereas the smallest model (CYBLE-416045-02) could be learned \new{in twelve minutes}. 
Even if the nRF52832 did not have the largest state space, the runtime was significantly higher compared to devices with the same state space (CC2650). The results presented in Table~\ref{tab:learning-results} show that learning the nRF52832 took more than five times as long as learning the CC2650. The difference in runtime occurred due to the extended waiting time for the nRF52832. This result indicates that the scalability of active automata learning did not only depend on the input alphabet size and state space of the \ac{sul}. Rather, we assume that the overhead to create a deterministic learning setup, e.g., repeating queries or waiting for answers, also influenced the efficiency of active automata learning. 

Conforming to the state space, the number of performed output queries and steps increased. For the devices where we considered the connection input, also the number of connection errors seemed to align with the complexity of the behavioral model. This observation emphasized our assumption that message loss regularly occurs. This justified the overhead of a decent error-handling procedure to ensure that the \ac{sul} is adequately reset to the initial state before the output query is executed. 

Figure~\ref{fig:learned-model-CC2650} shows the learned model of the CC2650 and Figure~\ref{fig:learned-model-nRF52832} of the nRF52832. To provide a clear and concise representation, we merged and simplified transitions. \new{The $\mathsf{+}$-symbol summarizes input and output labels, since depicting all labels for all nine considered inputs would make the models hardly readable.} The unmodified learned models of all \acp{soc} considered in this case study are available online~\cite{bleLearning}. 
The comparison between the learned models of the CC2650 (Figure~\ref{fig:learned-model-CC2650}) and the nRF52832 (Figure~\ref{fig:learned-model-nRF52832}) shows that even models with the same number of states describe different \ac{ble} protocol stack implementations. We highlighted in red for both models the transitions that show a different behavior on the input $\mathsf{length\_rsp}$. The nRF52832 responded to an unrequested length response only with a \ac{ble} data packet and then completely reset the connection procedure. Therefore, executing an unexpected length response on the nRF52832 led to the initial state akin to the performance of a scan request. The CC2650, instead, reacted to an unrequested length response with a response containing the packet $\mathsf{LL\_UNKNOWN\_RSP}$ and remained in the same state.


\begin{figure}
	\centering
	\input{CC2652R1-model.tex}
	\caption{Model learned of CC2652R1. For clarity, some transitions are not displayed. The complete model is available online~\cite{bleLearning}.}\label{fig:CC2652R1-model}
\end{figure}
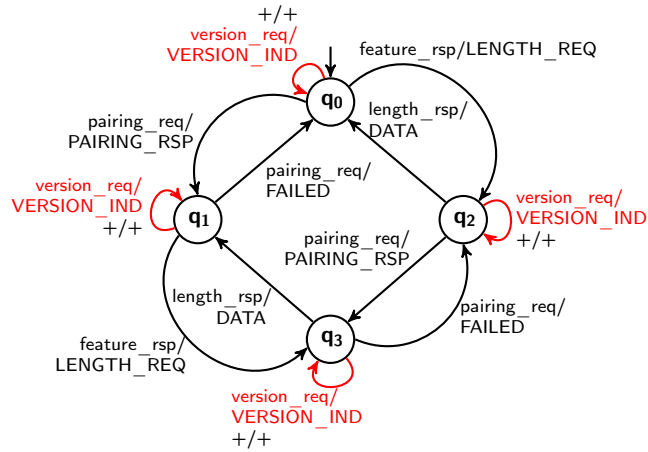

Figure~\ref{fig:CC2652R1-model} illustrates the learned model of the CC2652R1. Note that the initial state $q_0$ describes the behavior after the peripheral received a connection request. Therefore, the model defines the behavior of the parameter negotiation including the initiation of the pairing procedure. In our recent work \cite{DBLP:conf/nfm/PferscherA22}, we already reported that the learned model indicates inconsistency with the \ac{ble} specification. According to the \ac{ble} specification~\cite{Bluetooth53}, a version indication should be only answered once. As indicated by the red transitions, the CC2652R1 always responded with a version indication.



\begin{table}[t]
	\centering
	\caption{The non-deterministic behavior of the CC2640R2 \ac{ble} \ac{soc} disabled learning considering the entire input alphabet. The table shows the results of learning with a reduced input alphabet.}\label{tab:learning-results-cc2640r2}
	\resizebox{\textwidth}{!}{
	\input{learning-results-cc2640r2}}
\end{table}
Using the learning setup of Section~\ref{sec:learning-setup}, we could not learn the CC2640R2. Independent from the adaption of our error handling parameters, we always observed non-deterministic behavior. More interestingly, the non-deterministic behavior could repeatedly be observed on the following output query. 
{\small\[
\mathsf{connection\_req} \cdot \mathsf{pairing\_req} \cdot \mathsf{length\_rsp} \cdot \mathsf{length\_req} \cdot \mathsf{feature\_req}
\]}
In earlier stages of the learning procedure, we observed the following output sequence after the execution of the inputs.
{\scriptsize \[
\mathsf{LL\_LENGTH\_REQ} \cdot \mathsf{SM\_PAIRING\_RSP} \cdot \mathsf{BTLE\_DATA} \cdot \mathsf{LL\_LENGTH\_RSP} \cdot \underline{\mathsf{LL\_FEATURE\_RSP}}
\]}
Later in learning, we never again received any feature response for the input $\mathsf{feature\_req}$ if we executed this output query. The observed outputs always corresponded to the following sequence.
{\scriptsize
\[
\mathsf{LL\_LENGTH\_REQ} \cdot \mathsf{SM\_PAIRING\_RSP} \cdot \mathsf{BTLE\_DATA} \cdot \mathsf{LL\_LENGTH\_RSP} \cdot \underline{\mathsf{BTLE\_DATA}}
\]}

\new{We assume that the execution of the pairing request changed the internal behavior of the \ac{sul}. Hence, after the establishment of a certain number of pairing requests, the device failed to respond to provided requests.}
If we removed one of the inputs $\mathsf{pairing\_req}$, $\mathsf{length\_req}$, or $\mathsf{feature\_req}$, our learning setup successfully learned a deterministic model. Table~\ref{tab:learning-results-cc2640r2} shows the learning results for the CC2640R2 with the adapted input alphabets. Compared to the results in Table~\ref{tab:learning-results}, we observed more non-deterministic behavior than for the other devices, which led to repetitions of output queries. 

\subsubsection{Pairing Procedure Evaluation}

All experiments for learning the pairing procedure were performed on an HP EliteBook 840 G2 with an Intel i5-5200 operating at \unit[2.2]{GHz} and with \unit[16]{GB} RAM, running Ubuntu 20.04.2 LTS. We required a Linux-based operating system since we used the security manager interface provided by \textsc{SweynTooth}~\cite{sweyntooth}. This library creates valid field values for establishing encrypted communication. The provided module is implemented in C/C++ and uses the BlueZ library, which is a Bluetooth stack implementation for Linux. 



We slightly adapted the parameter configuration for learning the pairing procedure. Since we observed for the devices more non-deterministic behavior, we set $n_\mathrm{error} = 5$, $n_\mathrm{cache} = 3$, and the number of consecutive non-deterministic output queries to $n_\mathrm{nondet} = 3$. \new{ These numbers are lower than for learning the connection procedure, but in the case of a repeated connection error or non-deterministic error, we did not immediately abort the learning procedure.} Instead, the learning framework requested to hard reset the physical device. After the user performed a hard reset the learning procedure continued.

Table~\ref{tab:learning-results-pairing} presents the learning results of the pairing procedure. For this evaluation, we selected three out of the six devices, since these three devices supported the legacy pairing procedure and acted reasonably reliable. The learned models have between six and eleven states and the learning procedure took between \unit[0.9]{h} and \unit[5.2]{h}. Compared to the learning of the connection procedure, we observed significantly more non-deterministic outputs. We also extended Table~\ref{tab:learning-results-pairing} by the number of cached values that were changed during learning and the number of performed hard resets. 

\begin{table}[t]
	\centering
	\scriptsize
	\caption{Learning results of three out of six evaluated \ac{ble} \acp{soc}.}\label{tab:learning-results-pairing}
	\resizebox{\textwidth}{!}{
		\input{learning-results-pairing}}
\end{table}

The results show that the CC2640R2 and the CC2650 required hard resets, but the reasons for the hard reset were different. In the case of the CC2640R2, we observed that the \ac{soc} stopped accepting pairing requests after a certain amount of exchanged messages. Hence, repeated non-deterministic errors occur. More interestingly, the CC2650 required a hard reset since the device stopped responding to any request. The following sequence leads to the crash of the device:
\[
\mathsf{connection\_req} \cdot \mathsf{pairing\_req} \cdot \mathsf{confirm} \cdot \mathsf{random} \cdot
\] 
\vspace*{-0.6cm}
\[
\mathsf{encryption\_req} \cdot \langle\mathsf{pause\_encryption\_req}\rangle\cdot \mathsf{terminate\_ind}
\]
\begin{wrapfigure}[27]{l}{3.8cm}
	\vspace*{-1.2cm}
	\input{cyw43455-pairing-model.tex}
	\vspace*{-0.75cm}
	\caption{Model of CYW43455 pairing procedure. Labels are abbreviated and summarized. A complete model is available online~\cite{bleLearning}.}\label{fig:cyw43455-pairing-model}
\end{wrapfigure}
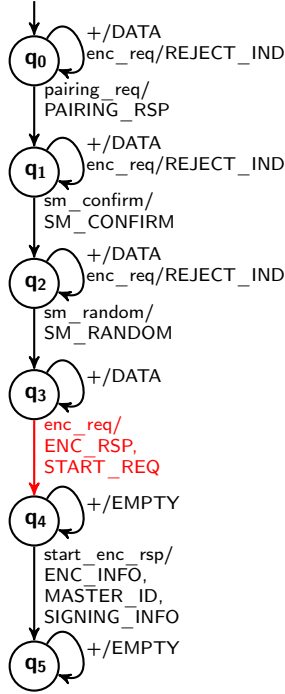
After the encryption request, the peripheral expected an encrypted response to the sent encrypted start request. However, our device sent a different unexpected encrypted message. After performing the reset, the CC2650 does not return to the advertising state. This shows that already the execution of an unexpected input sequence can trigger faulty behavior.

Figure~\ref{fig:cyw43455-pairing-model} presents the model of the pairing procedure of the CYW43455. We see that the message sequence conforms to the sequence chart shown in Figure~\ref{fig:ble-communication}. Encryption is enabled via the red transition between states $q_3$ and $q_4$. Afterward, only encrypted messages can be distributed. The self-loops for every state show that an unexpected input does not cancel the pairing procedure. 

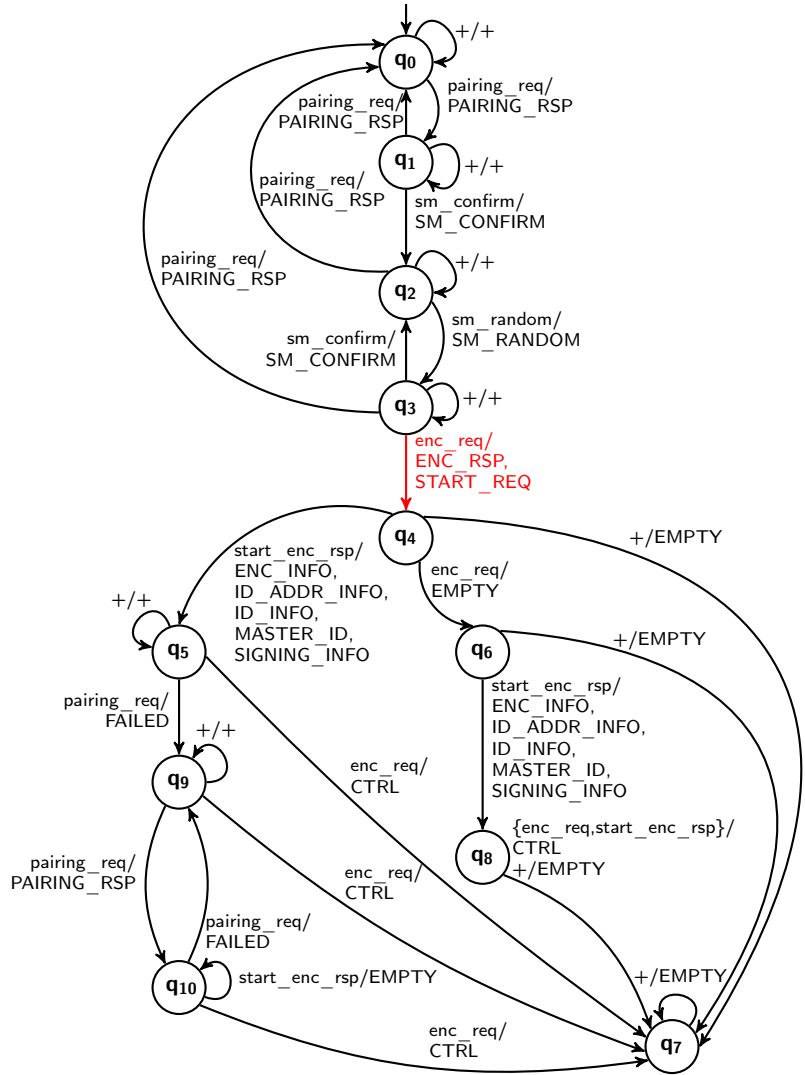
\begin{figure}[t]
	\centering
	\input{CC2640r2-pairing-model.tex}
	\caption{Model learned of CC2640R2 pairing procedure. For clarity, some transitions are not displayed. The complete model is available online~\cite{bleLearning}.}\label{fig:cc2640r2-pairing-model}
\end{figure}

Figure~\ref{fig:cc2640r2-pairing-model} describes the behavior of the CC2640R2. Compared to Figure~\ref{fig:cyw43455-pairing-model}, we see that unexpected inputs might revert the pairing procedure to previous states. The encryption is enabled after the transition from states  $q_3$ and $q_4$ is performed. In contrast to the behavior of CYW43455, we still could initiate further pairing procedures after keys have been already exchanged. Our presented learning results show that behavioral differences also occur during the pairing procedure and that the models also \new{enable to fingerprint} the tested devices.

\subsection{\ac{ble} Fingerprinting}

The comparison of the learned models of the connection procedure shows that all investigated \acp{soc} behave differently. Therefore, it is possible to fingerprint the \ac{soc}. The advantage of active automata learning, especially using $L^*$-based algorithms, is that every input is queried in each state to uniquely identify a state of the model. The collected query information can then be used to fingerprint the system. A closer look at the connection procedure models shows that even short input sequences sufficiently fingerprint the \ac{soc}.

\new{Even if random testing might be faster in the generation of a fingerprinting sequence for the investigated case study subjects. Still there exists many advantages in learning the behavioral model of the \ac{ble} device for fingerprinting. The learned behavioral models support the explainability of the found fingerprinting sequence. For example, the models illustrate in which state the models differ. An additional advantage of learning the model is that new fingerprinting sequences can be retrieved offline. For example, if we want to extend our set of investigated devices, we only learn the model of the new device. We can then check if the fingerprinting sequence is still valid to characterize uniquely all investigated devices. If not, we can use the learned models to generate a new fingerprinting sequence. This can be done offline, i.e., no active interaction with the \ac{ble} devices is required.}

\new{Lee and Yannakakis~\cite{DBLP:journals/pieee/LeeY96} discuss the conformance testing problem, also referred to as the fault-detection problem. To test the conformance between two systems, they define a so-called \emph{checking sequence} which is an input sequence that generates a different output sequence on both systems. In fingerprinting, we aim at generating such a checking sequence for a set of systems, where the output sequence should be unique for every device.}


\new{To generate such a checking sequence, we utilized the data structure that the learner generated during learning to build the hypothesis.}
Table~\ref{tab:fingerprinting} shows the observable outputs for each input after performing the initial connection request, i.e., the table shows the outputs that identify the state for the corresponding \ac{soc}. We determined that the set of observable outputs after an initial connection request is different for every \ac{soc}. 

\begin{table}[t]
	\centering
	\caption{The investigated \acp{soc} can be identified by only a single model state that is reached after performing an initial connection request. The columns of the table present the outputs that are observed when the input (row) is executed in the connection state.  The observable outputs show that only two inputs are required to distinguish the \acp{soc}.}\label{tab:fingerprinting}
	\resizebox{\textwidth}{!}{
		\input{fingerprinting}}
\end{table}


A closer look at the observable outputs shows that a combination of only two observable outputs is enough to identify the \ac{soc}. We highlight in Table~\ref{tab:fingerprinting} potential output combinations that depict the fingerprint of an \ac{soc}. We note that also other output combinations are possible. We can now use the corresponding inputs to generate a single output query that uniquely identifies one of our investigated \acp{soc}. Under the consideration that a scan request resets the \ac{soc}, we define the fingerprinting sequence for the \new{six} \acp{soc} output query as follows:
\[
\mathsf{scan\_req} \cdot \mathsf{connection\_req} \cdot \mathsf{feature\_rsp} \cdot \mathsf{scan\_req} \cdot \mathsf{connection\_req} \cdot \mathsf{version\_req} 
\]
The execution of this output query leads to a different observed output sequence for each of the five investigated \acp{soc}. For example, the corresponding output sequence for the nRF52832 is
\[
\mathsf{ADV} \cdot \mathsf{SM\_HDR} \cdot \mathsf{LL\_UNKNOWN\_RSP} \cdot \mathsf{ADV} \cdot \mathsf{BTLE\_DATA} \cdot \mathsf{LL\_VERSION\_IND},  
\]
whereas the sequence for the CC2650 is 
\[
\mathsf{ADV} \cdot \mathsf{BTLE\_DATA} \cdot \mathsf{BTLE\_DATA} \cdot \mathsf{ADV} \cdot \mathsf{BTLE\_DATA} \cdot \mathsf{LL\_VERSION\_IND}.
\]

The proposed manual analysis serves as a proof of concept that active automata learning can be used for fingerprinting \ac{ble} \acp{soc}. Obviously, the found input sequences for fingerprinting are only valid for the given \acp{soc}. For other \acp{soc}, a new model for every \ac{soc} should be learned to identify a possibly extended set of input sequences for fingerprinting. However, we recommend replacing the manual analysis with an automatic conformance testing technique between the models akin to Lee and Yannakakis~\cite{DBLP:journals/pieee/LeeY96} or Tappler et al.~\cite{DBLP:conf/icst/TapplerAB17}. 


%% file: devices.tex
\begin{tabular}{|l|l|l|}
	\hline
	\textbf{Manufacturer (Board)}  &  \textbf{SoC} & \textbf{Application}\\ \hline
	Texas Instruments  (LAUNCHXL-CC2640R2) & CC2640R2   & CC2640R2 LaunchPad\\
	Texas Instruments  (LAUNCHXL-CC2650) &  CC2650   & Project Zero\\
	Texas Instruments  (LAUNCHXL-CC26X2R1) &  CC2652R1   & Project Zero\\
	Cypress (CY8CPROTO-063-BLE) &  CYBLE-416045-02  & Find Me Target\\
	Cypress (Raspberry Pi 4 Model B) & CYW43455 & bluetoothctl \\
	Nordic (decaWave DWM1001-DEV) &  nRF52832  & Nordic GATTS\\
	\hline
\end{tabular}

%% file: learning-results.tex
\begin{tabular}{|l|P{1.3cm}|P{1.9cm}|P{1.65cm}|P{2cm}|c|}
	\hline
	 &  \textbf{CC2650} &  \textbf{CC2652R1}\textsuperscript{$\dagger$} &  \textbf{CYBLE-} \newline \textbf{416045-02}  & \textbf{CYW43455}\textsuperscript{$\dagger$} &  \textbf{nRF52832} \\ \hline
	\textbf{\# States} & 5 & 4  & 3   & 16 &  5\\ \hline
	\textbf{Total Time in minutes (min)} & 23.61 & 5.57 & 12.00   & 65.12 & 126.24 \\ 
	\textbf{Learning (min)} & 18.22 & 3.46 & 9.41  & 51.07& 73.80\\
	\textbf{Conformance Checking (min)} & 5.39 & 2.11 &  2.59  & 14.05& 52.44\\ \hline
	\textbf{\# Output Queries} & 405 & 196  & 243  & 784& 405 \\
	\textbf{\# Output Query Steps} & 1542  & 588 & 747   &  3136 &  1459\\ \hline
	\textbf{\# Conformance Tests} & 59 & 44	  & 32   &  164&  50\\
	\textbf{\# Conformance Test Steps} & 626 & 467 & 344   &  1958 &  580\\ \hline
	\textbf{\# Connection Errors} & 526 & - & 292   &  - &  459\\
	\textbf{\# Non-Deterministic Outputs} & 5  & 1 & 0   & 3 & 1 \\ \hline
\end{tabular}

%% file: automaton-cc2650.tex
\begin{tikzpicture}[>=stealth',font=\scriptsize\sffamily,thick] 
	\node[state,initial, align=center,initial text=,initial where=left,font=\footnotesize\bfseries\sffamily,text width=0.25cm,minimum size=1pt] (q0) {q\textsubscript{0}};
	\node[state, align=center,right = 2.5cm of q0,font=\footnotesize\bfseries\sffamily,text width=0.25cm,minimum size=1pt] (q1) {q\textsubscript{1}};
	\node[state, align=center,right = 3.5cm of q1,font=\footnotesize\bfseries\sffamily,text width=0.25cm,minimum size=1pt] (q4) {q\textsubscript{4}};
	\node[state,align=center,below = 1cm of $(q4)!0.5!(q1)$,font=\footnotesize\bfseries\sffamily,text width=0.25cm,minimum size=1pt] (q2) {q\textsubscript{2}};
	\node[state,font=\footnotesize\bfseries\sffamily,text width=0.25cm,minimum size=1pt,above = 1.1cm of $(q4)!0.5!(q1)$] (q3) {q\textsubscript{3}};

	\path[->] (q0.35) edge[bend left=45,looseness=0.9] node [above right=-0.05cm and -1.2cm,text width=2.5cm] {connect\_req/DATA} (q1.160)
	(q1.180) edge[] node [below right=-0.02cm and -0.75cm,text width=2.5cm] {scan/ADV} (q0.0)
	(q1.270) edge[bend right=20] node [below left=0cm and -0.25cm, align=right,text width=2.25cm] {pairing\_req/\\PAIRING\_RSP} (q2.180)
	(q2.130) edge[] node [above right=-0.25cm and 0.1cm, align=left,text width=2.5cm] {connect\_req/DATA\\pairing\_req/FAILED} (q1.330)
	(q2.0) edge[bend right=40] node [below right=-0.5cm and 0.35cm, align=left,text width=3cm] {version\_req/VERSION\_IND} (q4.270)
	(q3.0) edge[bend left=40] node [above left=0.1cm and -0.8cm, align=left,text width=1.75cm] {pairing\_req/\\PAIRING\_RSP} (q4.90)
	(q1.60) edge[bend left=0] node [above left=0.2cm and -0.5cm, align=right,text width=2cm] {version\_req/\\VERSION\_IND} (q3.200)
	(q3.270) edge[bend left=40] node [above right=-0.1cm and 0.45cm, align=left,text width=1.75cm] {connect\_req/\\DATA} (q1.40)
	(q4.180) edge[bend right=0] node [above right = -0.1cm and -0.75cm, align=left,text width=2.5cm] {connect\_req/DATA} (q1.0)
	(q3.150) edge[bend right=40,] node [above right = 0.05cm and -0.45cm, align=left,text width=2.5cm] {scan/ADV} (q0.100)
	(q4.20) edge[bend right=100,looseness=1.3] node [above right = 0cm and -0.25cm, align=left,text width=2.5cm] {scan/ADV} (q0.140)
	(q0) edge[loop,in=200,out=260,looseness=4] node [below=0cm,text width=1.5cm] {scan/ADV\\+/EMPTY} (q0)
	(q2.240) edge[bend left=40] node [below left = -0.5cm and 1cm, align=left,text width=2cm, align=right] {scan/ADV} (q0.270)
	(q1) edge[loop,in=140,out=70,looseness=4] node [above right =-0.1cm and 0cm,text width=1cm] {+/+} (q1)
	(q1) edge[loop,in=190,out=260,looseness=4,red] node [below left =0cm and -0.5cm,text width=2.75cm] {length\_rsp/UNKNOWN} (q1)
	(q2) edge[loop,in=270,out=340,looseness=4,red] node [right=0.1cm,text width=2.75cm] {length\_rsp/UNKNOWN\\+/+} (q2)
	(q3) edge[loop,in=290,out=350,looseness=4] node [right=0cm,text width=1cm] {+/+} (q3)
	(q3) edge[loop,in=130,out=60,looseness=4,red] node [above left=-0.05cm and -1cm,text width=2.75cm,align=right] {length\_rsp/UNKNOWN} (q3)
	(q4) edge[loop,in=0,out=290,looseness=4,red] node [above right=-0.25cm and 0cm,text width=2.75cm] {\mbox{+/+}\\\mbox{length\_rsp/UNKNOWN}} (q4)
	;
\end{tikzpicture}

%% file: automaton-nrf52832.tex
\begin{tikzpicture}[>=stealth',font=\scriptsize\sffamily,thick] 
	\node[state,initial, align=center,initial text=,initial where=left,font=\footnotesize\bfseries\sffamily,text width=0.25cm,minimum size=1pt] (q0) {q\textsubscript{0}};
	\node[state, align=center,right = 2.5cm of q0,font=\footnotesize\bfseries\sffamily,text width=0.25cm,minimum size=1pt] (q1) {q\textsubscript{1}};
	\node[state, align=center,right = 3.5cm of q1,font=\footnotesize\bfseries\sffamily,text width=0.25cm,minimum size=1pt] (q4) {q\textsubscript{4}};
	\node[state,align=center,below = 1cm of $(q4)!0.5!(q1)$,font=\footnotesize\bfseries\sffamily,text width=0.25cm,minimum size=1pt] (q2) {q\textsubscript{2}};
	\node[state,font=\footnotesize\bfseries\sffamily,text width=0.25cm,minimum size=1pt,above = 1.1cm of $(q4)!0.5!(q1)$] (q3) {q\textsubscript{3}};
	
	\path[->] (q0.35) edge[bend left=45,looseness=0.9] node [above right=-0.05cm and -1.2cm,text width=2.5cm] {connect\_req/SM\_RSP} (q1.160)
	(q1.180) edge[red] node [below right=-0.02cm and -0.75cm,text width=2.5cm] {\textcolor{black}{scan/ADV}\\length\_rsp/DATA} (q0.0)
	(q1.60) edge[bend left=0] node [above left=0.1cm and -0.5cm, align=right,text width=1.8cm] {version\_req/\\VERSION\_IND} (q3.200)
	(q1.270) edge[bend right=0] node [below left=0cm and -0.25cm, align=right,text width=2.5cm] {mtu\_req/MTU\_RSP} (q2.180)
	(q2.0) edge[bend right=40] node [below right=-0.5cm and 0.35cm, align=left,text width=3cm] {version\_req/VERSION\_IND} (q4.270)
	(q3.0) edge[bend left=40] node [above left=0.2cm and -1cm, align=left,text width=1.75cm] {mtu\_req/\\MTU\_RSP} (q4.90)
	(q3.270) edge[bend left=40] node [above right=-0.1cm and 0.45cm, align=left,text width=1.75cm] {connect\_req/\\SM\_RSP} (q1.40)
	(q2.100) edge[bend right=40] node [above right=-0.7cm and 0.55cm, align=left,text width=2cm] {connect\_req/\\SM\_RSP} (q1.320)
	(q4.180) edge[bend right=0] node [above right = -0.1cm and -1cm, align=left,text width=2.75cm] {connect\_req/SM\_RSP} (q1.0)
	(q3.150) edge[bend right=40,red] node [above right = 0.05cm and -0.25cm, align=left,text width=2.5cm] {\textcolor{black}{scan/ADV}\\length\_rsp/DATA} (q0.100)
	(q4.20) edge[bend right=100,looseness=1.3,red] node [above right = 0cm and -0.25cm, align=left,text width=2.5cm] {\textcolor{black}{scan/ADV}\\length\_rsp/DATA} (q0.140)
	(q2.240) edge[bend left=40,red] node [below left = -0.5cm and 1cm, align=left,text width=2cm, align=right] {\textcolor{black}{scan/ADV}\\length\_rsp/DATA} (q0.270)
	(q0) edge[loop,in=200,out=260,looseness=4] node [below=0cm,text width=1.5cm] {scan/ADV\\+/EMPTY} (q0)
	(q1) edge[loop,in=140,out=70,looseness=4] node [above right =-0.1cm and 0cm,text width=1cm] {+/+} (q1)
	(q2) edge[loop,in=270,out=340,looseness=4] node [right=0.1cm,text width=2.25cm] {mtu\_req/MTU\_ERR\\+/+} (q2)
	(q3) edge[loop,in=290,out=350,looseness=4] node [right=0cm,text width=1cm] {+/+} (q3)
	(q3) edge[loop,in=130,out=60,looseness=4] node [above left=-0.05cm and -1cm,text width=2cm,align=center] {version\_req/\\DATA} (q3)
	(q4) edge[loop,in=0,out=290,looseness=4] node [above right=-0.25cm and 0cm,text width=2.75cm] {\mbox{+/+}\\\mbox{version\_req/DATA} mtu\_req/MTU\_ERR} (q4)
	;
\end{tikzpicture}

%% file: learning-results-cc2640R2.tex
\begin{tabular}{|l|c|c|c|}
	\hline
	 & \textbf{no $\mathsf{pairing\_req}$} & \textbf{no $\mathsf{length\_req}$}  & \textbf{no $\mathsf{feature\_req}$} \\ \hline
	\textbf{\# States} & 6 & 11  & 11  \\ \hline
	\textbf{Total Time (min)} & 26.40 & 47.57 & 40.29   \\ 
	\textbf{Learning Time (min)} &  16.94 & 30.73 & 28.29   \\
	\textbf{Conformance Checking Time (min)} & 9.46 & 16.84 &  11.70\\ \hline
	\textbf{\# Output Queries} & 384 &  705 & 704  \\
	\textbf{\# Output Query Steps} & 1474  & 3143 & 3143  \\ \hline
	\textbf{\# Conformance Tests} & 61 & 115  & 111   \\
	\textbf{\# Conformance Test Steps} & 712 & 1406 & 1371   \\ \hline
	\textbf{\# Connection Errors} & 449 & 822 & 821  \\
	\textbf{\# Non-Deterministic Outputs} & 1  & 10 & 2  \\ \hline
\end{tabular}

%% file: learning-results-pairing.tex
\begin{tabular}{|l|c|c|c|}
	\hline
	 & \textbf{CC2640R2} & \textbf{CC2650}  & \textbf{CYW43455} \\ \hline
	\textbf{\# States} & 11 & 10  & 6  \\ \hline
	\textbf{Total Time (min)} &133.01 & 312.37 & 52.72   \\ 
	\textbf{Learning Time (min)} &  116.95 & 201.34 & 38.83   \\
	\textbf{Conformance Checking Time (min)} & 16.06 & 111.03 &  13.89\\ \hline
	\textbf{\# Output Queries} & 487 &  453 & 223  \\
	\textbf{\# Output Query Steps} & 3142  & 2869 & 1012  \\ \hline
	\textbf{\# Conformance Tests} & 110 & 100  & 60   \\
	\textbf{\# Conformance Test Steps} & 273 & 601 & 287   \\ \hline
	\textbf{\# Non-Deterministic Outputs} & 133  & 80 & 29  \\ 
	\textbf{\# Cache Updates} & 1 & 3 & 0  \\
	\textbf{\# Hard Resets} & 6 &11 & 0  \\\hline
\end{tabular}

%% file: cyw43455-pairing-model.tex
\begin{tikzpicture}[>=stealth',font=\scriptsize\sffamily,thick,text width=0.4cm] 
	\node[state,initial,initial text=,initial where=above,align=center,font=\footnotesize\bfseries\sffamily,text width=0.25cm,minimum size=1pt] (q0) {q\textsubscript{0}};
	\node[state,below = 0.8cm of q0,align=center,font=\footnotesize\bfseries\sffamily,text width=0.25cm,minimum size=1pt] (q1) {q\textsubscript{1}};
	\node[state,below = 0.8cm of q1,align=center,font=\footnotesize\bfseries\sffamily,text width=0.25cm,minimum size=1pt] (q2) {q\textsubscript{2}};
	\node[state,below = 0.8cm of q2,align=center,font=\footnotesize\bfseries\sffamily,text width=0.25cm,minimum size=1pt] (q3) {q\textsubscript{3}};
	\node[state,below = 1cm of q3,align=center,font=\footnotesize\bfseries\sffamily,text width=0.25cm,minimum size=1pt] (q4) {q\textsubscript{4}};
	\node[state,below = 1.25cm of q4,align=center,font=\footnotesize\bfseries\sffamily,text width=0.25cm,minimum size=1pt] (q5) {q\textsubscript{5}};

	\path[->]

	(q0.270) edge[] node [above right = -0.2cm and 0cm, align = left,text width=1.75cm] {pairing\_req/\\PAIRING\_RSP} (q1.90)
	(q1.270) edge[] node [above right = -0.2cm and 0cm, align = left,text width=1.75cm] {sm\_confirm/\\SM\_CONFIRM} (q2.90)
	(q2.270) edge[] node [above right = -0.2cm and 0cm, align = left,text width=1.75cm] {sm\_random/\\SM\_RANDOM} (q3.90)
	(q3.270) edge[red] node [above right = -0.4cm and 0cm, align = left,text width=1.75cm] {enc\_req/\\ENC\_RSP,\\START\_REQ} (q4.90)
	(q4.270) edge[] node [above right = -0.6cm and 0cm, align = left,text width=1.75cm] {start\_enc\_rsp/\\ENC\_INFO,\\MASTER\_ID,\\SIGNING\_INFO} (q5.90)
	(q0) edge[loop,in=340,out=60,looseness=4] node [right=-0.05cm,text width=2.5cm,align=left] {+/DATA\\enc\_req/REJECT\_IND} (q0)
	(q1) edge[loop,in=340,out=60,looseness=4] node [right=-0.05cm,text width=2.5cm,align=left] {+/DATA\\enc\_req/REJECT\_IND} (q1)
	(q2) edge[loop,in=340,out=60,looseness=4] node [right=-0.05cm,text width=2.5cm,align=left] {+/DATA\\enc\_req/REJECT\_IND} (q2)
	(q3) edge[loop,in=340,out=60,looseness=4] node [right=-0.05cm,text width=2.5cm,align=left] {+/DATA} (q3)
	(q4) edge[loop,in=340,out=60,looseness=4] node [right=-0.05cm,text width=2.5cm,align=left] {+/EMPTY} (q4)
	(q5) edge[loop,in=340,out=60,looseness=4] node [right=-0.05cm,text width=2.5cm,align=left] {+/EMPTY} (q5)
	;
\end{tikzpicture}

%% file: CC2640r2-pairing-model.tex
\begin{tikzpicture}[>=stealth',font=\scriptsize\sffamily,thick,text width=0.4cm] 
	\node[state,initial,initial text=,initial where=above,align=center,font=\footnotesize\bfseries\sffamily,text width=0.35cm,minimum size=1pt] (q0) {q\textsubscript{0}};
	\node[state,below = 0.6cm of q0,align=center,font=\footnotesize\bfseries\sffamily,text width=0.35cm,minimum size=1pt] (q1) {q\textsubscript{1}};
	\node[state,below = 1cm of q1,align=center,font=\footnotesize\bfseries\sffamily,text width=0.35cm,minimum size=1pt] (q2) {q\textsubscript{2}};
	\node[state,below = 0.8cm of q2,align=center,font=\footnotesize\bfseries\sffamily,text width=0.35cm,minimum size=1pt] (q3) {q\textsubscript{3}};
	\node[state,below = 1cm of q3,align=center,font=\footnotesize\bfseries\sffamily,text width=0.35cm,minimum size=1pt] (q4) {q\textsubscript{4}};
	\node[state,below left = 1cm and 2.5cm of q4,align=center,font=\footnotesize\bfseries\sffamily,text width=0.35cm,minimum size=1pt] (q5) {q\textsubscript{5}};
	\node[state,below right= 1cm and 0.5cm of q4,align=center,font=\footnotesize\bfseries\sffamily,text width=0.35cm,minimum size=1pt] (q6) {q\textsubscript{6}};
	\node[state,below = 2cm of q6,align=center,font=\footnotesize\bfseries\sffamily,text width=0.35cm,minimum size=1pt] (q8) {q\textsubscript{8}};
	\node[state,below = 1cm of q5,align=center,font=\footnotesize\bfseries\sffamily,text width=0.35cm,minimum size=1pt] (q9) {q\textsubscript{9}};
	\node[state,below = 2cm of q9,align=center,font=\footnotesize\bfseries\sffamily,text width=0.35cm,minimum size=1pt] (q10) {q\textsubscript{10}};
	\node[state,below right = 2cm and 2cm of q8,align=center,font=\footnotesize\bfseries\sffamily,text width=0.35cm,minimum size=1pt] (q7) {q\textsubscript{7}};

	\path[->]

	(q0.320) edge[bend left = 45] node [above right = -0.2cm and 0cm, align = left,text width=1.75cm] {pairing\_req/\\PAIRING\_RSP} (q1.50)
	(q1.90) edge[] node [above left = -0.4cm and -0.1cm, align = right,text width=1.75cm] {pairing\_req/\\PAIRING\_RSP} (q0.270)
	(q2.130) edge[bend left = 90, looseness = 2.2] node [above right = -0.6cm and -0.02cm, align = left,text width=1.75cm] {pairing\_req/\\PAIRING\_RSP} (q0.190)
	(q3.190) edge[bend left = 90, looseness = 2.2] node [above right = -1cm and 0.1cm, align = left,text width=1.75cm] {pairing\_req/\\PAIRING\_RSP} (q0.140)
	(q1.270) edge[] node [above right = -0.2cm and 0cm, align = left,text width=1.75cm] {sm\_confirm/\\SM\_CONFIRM} (q2.90)
	(q2.340) edge[bend left = 45] node [above right = -0.2cm and 0cm, align = left,text width=1.75cm] {sm\_random/\\SM\_RANDOM} (q3.60)
	(q3.90) edge[] node [above left = -0.4cm and 0cm, align = right,text width=1.75cm] {sm\_confirm/\\SM\_CONFIRM} (q2.270)
	(q3.270) edge[red] node [above right = -0.4cm and 0cm, align = left,text width=1.75cm] {enc\_req/\\ENC\_RSP,\\START\_REQ} (q4.90)
	(q4.120) edge[bend right = 45] node [below right = 0.1cm and -0.5cm, align = left,text width=2.25cm] {start\_enc\_rsp/\\ENC\_INFO,\\ID\_ADDR\_INFO,\\ID\_INFO,\\MASTER\_ID,\\SIGNING\_INFO} (q5.90)
	(q4.300) edge[bend right = 45] node [below right = -0.7cm and -0.15cm, align = left,text width=2cm] {enc\_req/\\EMPTY} (q6.110)
	(q4.50) edge[bend left = 60, looseness=1.6] node [above left = 1.5cm and -0.15cm, align = left,text width=2cm] {+/EMPTY} (q7.0)
	(q6.50) edge[bend left = 60, looseness=1.6] node [above left = 1.25cm and -0.15cm, align = left,text width=2cm] {+/EMPTY} (q7.30)
	(q6.270) edge[] node [below right = -1.15cm and 0cm, align = left,text width=2.25cm] {start\_enc\_rsp/\\ENC\_INFO,\\ID\_ADDR\_INFO,\\ID\_INFO,\\MASTER\_ID,\\SIGNING\_INFO} (q8.90)
	(q8.320) edge[bend left = 25] node [above left = 0.6cm and -3cm, align = left,text width=4cm] {\{enc\_req,start\_enc\_rsp\}/\\CTRL\\+/EMPTY } (q7.140)
	(q5.350) edge[bend right = 5] node [above right = 0.75cm and -1cm, align = left,text width=4cm] {enc\_req/\\CTRL} (q7.160)
	(q9.330) edge[bend right = 10] node [above right = 0.5cm and -1cm, align = left,text width=4cm] {enc\_req/\\CTRL} (q7.190)
	(q5.270) edge[bend right = 0] node [above left = -0.25cm and 0cm, align = right,text width=1.75cm] {pairing\_req/\\FAILED} (q9.90)
	(q9.240) edge[bend right = 25] node [above left = -0.25cm and 0cm, align = right,text width=1.75cm] {pairing\_req/\\PAIRING\_RSP } (q10.120)
	(q10.70) edge[bend right = 25] node [below right = 0.25cm and -0.15cm, align = left,text width=1.75cm] {pairing\_req/\\FAILED} (q9.290)
	(q10.320) edge[bend right = 15] node [above right = 0cm and 0cm, align = left,text width=1.75cm] {enc\_req/\\CTRL} (q7.210)

	(q0) edge[loop,in=0,out=70,looseness=4] node [right=-0.05cm,text width=2.5cm,align=left] {+/+} (q0)
	(q1) edge[loop,in=320,out=30,looseness=4] node [right=-0.05cm,text width=2.5cm,align=left] {+/+} (q1)
	(q2) edge[loop,in=360,out=70,looseness=4] node [right=-0.05cm,text width=2.5cm,align=left] {+/+} (q2)
	(q3) edge[loop,in=340,out=40,looseness=4] node [right=-0.05cm,text width=2.5cm,align=left] {+/+} (q3)
	(q5) edge[loop,in=170,out=110,looseness=4] node [above left=0cm and -0.75cm,text width=1cm,align=left] {+/+} (q5)
	(q9) edge[loop,in=60,out=0,looseness=4] node [above left=0.1cm and -0.75cm,text width=1cm,align=left] {+/+} (q9)
	(q10) edge[loop,in=40,out=340,looseness=4] node [right=0cm,text width=3cm,align=left] {start\_enc\_rsp/EMPTY} (q10)
	(q7) edge[loop,in=110,out=50,looseness=4] node [above right=0cm and -0.75cm,text width=3cm,align=left] {+/EMPTY} (q7)
	;
\end{tikzpicture}

%% file: fingerprinting.tex
\begin{tabular}{|l|c|c|c|c|}
	\hline
	& \textbf{feature\_rsp} & \textbf{version\_req} & \textbf{length\_req}  & \textbf{length\_rsp}  \\ \hline
	CC2640R2 & BTLE\_DATA & \textbf{BTLE\_DATA} & \textbf{LL\_LENGTH\_RSP} & \textbf{BTLE\_DATA}\\ 
	CC2650 & BTLE\_DATA & \textbf{LL\_VERSION\_IND} & \textbf{LL\_UNKNOWN\_RSP} & \textbf{LL\_UNKNOWN\_RSP}\\ 
	CC2652R1 & \textbf{LL\_LENGTH\_REQ }& LL\_VERSION\_IND  & LL\_LENGTH\_RS & BTLE\_DATA\\
	CYBLE-416045-02 &\textbf{ LL\_REJECT\_IND} & LL\_VERSION\_IND  & LL\_UNKNOWN\_RSP & LL\_UNKNOWN\_RSP\\ 
	CYW43455 & \textbf{ATT\_MTU\_REQ} & LL\_VERSION\_IND  & LL\_LENGTH\_RS & LL\_REJECT\_IND \\ 
	nRF52832 &\textbf{LL\_UNKNOWN\_RSP} & LL\_VERSION\_IND  & LL\_LENGTH\_RS & BTLE\_DATA\\ \hline
\end{tabular}

%% file: related-work.tex

Celosia and Cunche~\cite{DBLP:conf/ccs/CelosiaC19} also investigated fingerprinting \ac{ble} devices, however, their proposed methodology is based on the \ac{gatt}, whereas our technique also operates on different layers, e.g., the \ac{ll} or \ac{sm}, of the \ac{ble} protocol stack. Their proposed fingerprinting method is based on a large dataset containing information that can be obtained from the \ac{gatt} profile, like services and characteristics. 

Argyros et al.~\cite{DBLP:conf/ccs/ArgyrosSJKK16} discuss the combination of active automata learning and differential testing to fingerprint the \acp{sul}. They propose a framework where they first learn symbolic finite automata of different implementations and then automatically analyze differences between the learned models. They evaluated their technique on implementations of TCP, web application firewalls, and web browsers. A similar technique was proposed by Tappler et al.~\cite{DBLP:conf/icst/TapplerAB17} investigating the \ac{mqtt} protocol. However, their motivation was not to fingerprint \ac{mqtt} brokers, but rather test for inconsistencies between the learned models. These found inconsistencies show discrepancies to the \ac{mqtt} specification. Following an akin idea, but motivated by security testing, several communication protocols like TLS \cite{DBLP:conf/uss/RuiterP15}, TCP \cite{DBLP:conf/cav/Fiterau-Brostean16}, SSH \cite{DBLP:conf/spin/Fiterau-Brostean17} or DTLS \cite{DBLP:conf/uss/Fiterau-Brostean20} have been learning-based tested. In the literature, these techniques are denoted as protocol state fuzzing. To the best of our knowledge, none of these techniques interacted with an implementation on an external physical device, but rather interacted via localhost or virtual connections with the \acp{sul}.

One protocol state fuzzing technique on physical devices was proposed by Stone et al.~\cite{DBLP:conf/esorics/StoneCR18}. They detected security vulnerabilities in the 802.11 4-Way handshake protocol by testing Wi-Fi routers. Aichernig et al.~\cite{DBLP:conf/nfm/AichernigBK19} propose an industrial application for learning-based testing of measurement devices in the automotive industry. Both case studies emphasize our observation that non-deterministic behavior hampers the inference of behavioral models via active automata learning. Other physical devices that have been learned are bank cards \cite{DBLP:conf/icst/AartsRP13} and biometric passports \cite{DBLP:conf/isola/AartsSV10}. The proposed techniques use a USB-connected smart card reader to interact with the cards. Furthermore, Chalupar et al.~\cite{DBLP:conf/woot/ChaluparPPR14} used Lego\textsuperscript{\textregistered} to create an interface to learn the model of a smart card reader. \new{In the context of protocol fuzzing, we showed in a follow-up of this article that the learned \ac{ble} models can be used to reveal robustness issues in \ac{ble} devices.}

%% file: conclusion.tex
\subsection{Summary}
In this article, we presented a case study on learning-based testing of the \ac{ble} protocol. The case study aimed to evaluate learning-based testing in a practical setup. For this, we proposed a general learning architecture for \ac{ble} devices. The proposed architecture enabled the inference of a model that describes the behavior of a \ac{ble} protocol implementation.  We evaluated our presented learning framework in a case study consisting of six \ac{ble} devices. The results of the case study showed that the active learning of a behavioral model is possible in a practicable amount of time. However, our evaluation showed that \new{extensions} to state-of-the-art learning algorithms, such as including error-handling procedures, were required for successful model inference. By using learning-based testing, we revealed that one device crashes on the execution of an in-depth testing sequence. Furthermore, the learned models depicted that implementations of the \ac{ble} stack vary significantly from device to device. This observation confirmed our hypothesis that active automata learning enables fingerprinting of black-box systems.

\subsection{Discussion}
We successfully applied active automata learning to reverse engineer the behavioral models of \ac{ble} devices. \new{We experienced challenges in creating a reliable and general learning framework to learn a behavioral model of a wireless protocol implemented on a physical device. To learn deterministic models, we needed to repeat executions on the \ac{sul}. Especially, the required guarantee of a reliable reset created issues. Another possibility to overcome this issue would have been to use a resetless learning algorithm as proposed by Rivest and Schapire~\cite{DBLP:journals/iandc/RivestS93}. However, packet loss and delayed responses still present a problem in such algorithms, since we might assume that we are in the wrong state. Hence, these algorithms would also require countermeasures against non-deterministic observations.}
The advantage is that \ac{ble} interface creation only needs to be done once. Our proposed framework, which is also publicly available~\cite{bleLearning}, can now be used for learning the behavioral models of many \ac{ble} devices. Our presented learning results show that in practice the scalability of active automata learning not only depended on the efficiency of the underlying learning algorithm but also on the overhead due to \ac{sul} interaction. However, once this interface was created, active automata learning successfully revealed a robustness issue in a tested \ac{ble} device.
All of the learned models show behavioral differences in the \ac{ble} protocol stack implementations. Therefore, we can use active automata learning to fingerprint the underlying \ac{soc} of a black-box \ac{ble} device. The possibility to fingerprint the \ac{ble} could be a possible security issue since it enables an attacker to exploit specific vulnerabilities, e.g., from a \ac{ble} vulnerability collection like \textsc{SweynTooth} \cite{DBLP:conf/usenix/Garbelini00SK20}. 
Compared to the \ac{ble} fingerprinting technique of Celosia and Cunche \cite{DBLP:conf/ccs/CelosiaC19}, our proposed technique is data and time-efficient. Instead of collecting  $13\,000$ data records over five months, we can learn the models within hours.

\subsection{Future Work}
Our first investigation of the security-critical behavior of \ac{ble} devices could successfully reveal robustness issues via active automata learning.
Despite the found robustness issue, the learned models do not show any security vulnerabilities. However, for future work, we plan to consider additional functionality of the \ac{ble} protocol stack, e.g., the secure pairing procedure. Considering the public/private key-exchange procedure might reveal further security issues. 

Our proposed method was inspired by the work of Garbelini et al.~\cite{DBLP:conf/usenix/Garbelini00SK20} since their presented fuzz-testing technique demonstrated that model-based testing can be applied to \ac{ble} devices. Instead of creating the model manually, we showed that learning a behavioral model of the \ac{ble} protocol implemented on a physical device is possible. In recent work \cite{DBLP:conf/nfm/PferscherA22} we extended our proposed learning framework for learning-based fuzzing of the \ac{ble} protocol. For this, we used our learned models of the connection procedure to generate test cases for fuzzing. Our proposed technique successfully revealed several issues in the implementation of the \ac{ble} protocol. We are currently working on extending this technique for learning-based fuzzing the \ac{ble} pairing procedure.

We find that the non-deterministic behavior of the \ac{ble} devices hampered the learning of deterministic models. Instead of workarounds to overcome non-deterministic behavior, we could learn a non-deterministic model. We already applied non-deterministic learning to the \ac{mqtt}  protocol \cite{DBLP:conf/pts/PferscherA20}.  Following a similar idea, we could learn a non-deterministic model of the \ac{ble} protocol.